\documentclass[useAMS,usenatbib]{mn2e}
\usepackage{tipa}
\usepackage{amssymb}
\usepackage{times}
\usepackage{graphicx}
\usepackage{epsfig}
\usepackage{color}
\usepackage{booktabs}
\usepackage{txfonts}
\usepackage{pifont}

\newcommand{\Hb}{H$\beta$}
\newcommand{\Ha}{H$\alpha$}
\newcommand{\Hd}{H$\delta$}
\newcommand{\apj}{ApJ}
\newcommand{\apjs}{ApJS}
\newcommand{\apjl}{ApJL}
\newcommand{\aj}{AJ}
\newcommand{\mnras}{MNRAS}
\newcommand{\pasp}{PASP}
\newcommand{\pasj}{PASJ}
\newcommand{\aap}{A\&A}

\title[Recent Star-Forming Activity in Local Elliptical Galaxies]{Recent Star-Forming Activity in Local Elliptical Galaxies}
\author[Huang \& Gu]{Song Huang \thanks{E-mail:
song.clearskies@gmail.com}, Q.-S. Gu \thanks{E-mail: qsgu@nju.edu.cn}\\
 Department of Astronomy, Nanjing University, Nanjing, 210093, P.
R. China}

\begin{document}


\pagerange{\pageref{firstpage}--\pageref{lastpage}} \pubyear{2008}

\maketitle

\label{firstpage}

\begin{abstract}
\\
The formation and evolution of elliptical galaxies (EGs) is still an open
question. In particular, recent observations suggest that elliptical galaxies are
not only simple spheroidal systems of old stars. In this paper
we analyze a sample of  elliptical galaxies selected from the Sloan
Digital Sky Survey in order to study the star-forming activity in
local elliptical galaxies. Among these 487 ellipticals we find that
13 EGs show unambiguous evidence of recent star-formation activity betrayed by
conspicuous nebular emission lines. Using the evolutionary
stellar population synthesis models and Lick absorption line indices we derive stellar ages,
metallicities,   and $\alpha$-element abundances, and thus  reconstruct
the star formation and chemical evolution history of the star-forming elliptical galaxies (SFEGs) in our sample.

We find that SFEGs have relative younger  stellar population age, higher metallicity, and lower stellar mass,
 and that their star formation history can be well described by a recent minor and short starburst superimposed on old stellar component. We
also detect 11 E+A galaxies whose stellar population properties are closer to those of quiescent (normal) ellipticals than to star-forming ones. However, from the analysis of their
absorption line indices, we note that our E+A galaxies show a significant fraction of intermediate-age stellar populations, remarkably  different  from the quiescent galaxies. This
might suggest an evolutionary link between E+As and star-forming ellipticals.  Finally, we confirm the relations between age, metallicity, $\alpha$ element abundance, and stellar mass for local elliptical galaxies.
 \end{abstract}


\section{Introduction}

The formation and evolution of elliptical galaxies (EGs) is very
important for understanding galaxies formation and evolution, but
several key issues are still open questions. Early studies regarded
EGs as a family of galaxies that has very simple properties: smooth
morphology; old stellar population; red optical color; free of cold
gas and young star formation (Searle, Sargent \& Bagnuolo 1973;
Larson 1975). However, recent observations suggest that this
viewpoint is really oversimple for EGs. Astronomers already detected
cold gas and dust and even recent/residual star formation in EGs
(Morganti et al. 2006; Yi et al. 2005; Kaviraj et al. 2007; see also
the recent review by Sarzi et al. 2008). Fukugita et al. (2004)
reported four star-forming EGs from Sloan Digital Sky Survey (SDSS)
DR2, where the star formation rate can be comparable with normal
spiral galaxies. With a much larger sample of 16,000 early-type
galaxies, Schawinski et al. (2007) found that about $4\%$ early type
galaxies show ongoing star formation, this fraction is based on
emission lines, but the fraction of these active elliptical or
early-type galaxies is actually highly dependent on the data and the
galaxy mass (see Schawinski et.al. 2006, 2007b). For example, with
NUV data from GALEX, this fraction could be as high as 30\% (Kaviraj
et.al. 2007).

The detection of  active star-forming EGs  provides a severe
challenge for galaxy formation and evolution model.  There are two
main competing  formation scenarios for EGs: the \emph{monolithic
collapse} model and the \emph{hierarchical merger} model. In the
early monolithic collapse model, EGs formed from the violent
starburst happened at very high redshift and evolved passively ever
since then (Eggen, Lynden-Bell \& Sandage 1962; Larson 1974; Arimoto
\& Yoshii 1987; Bressan, Chiosi \& Fagotto 1994). On the other hand,
the hierarchical merger model considers a more extended formation
history. In this model, the EGs are formed from the major merger of
disc galaxies of similar mass (Toomre 1977; White \& Rees 1978;
Kauffmann, White \& Guiderdoni 1993). Both these models can explain
some aspects of observed data and face some unavoidable problems at
the same time: the \emph{monolithic collapse} model is supported by
the existence and the tightness of scaling relation like
color-magnitude relation (CMR)(Sandage \& Visvanathan 1978),
fundamental plane (Djorgovski \& Davis 1987) and Mg2-$\sigma$
relation (Colless et al. 1999, Kuntschner et al. 2001) in EGs, but
can not explain the active star-formation and remarkable cold gas
found in EGs. For the \emph{hierarchical merger} model, now there
are lots of observational evidence that demonstrates the importance
of interaction and merger in the formation of early-type galaxies
and it is expected by the most popular $\Lambda$-CDM cosmology. More
and more recent observations indicate that some field EGs could be
formed at relative low redshift and the stellar populations can
spread widely, which also support the \emph{hierarchical merger}
model. But the observed "Down-Sizing" effect (Kodama et al. 2004;
Cimatti, Daddi \& Renzini 2006) is not consistent with the
\emph{hierarchical merger} model, which predicts that the most
massive EGs should be formed most recently. Although from recent
simulations, this "down-sizing" behaviour also could be obtained
within the hierarchical model (De Lucia et.al. 2006). These two
models predict totally different star formation history (SFH) for
EGs, and the recent star formation activity can be an important
tracer of SFH in EGs. The time-scale and intensity of the recent
star-formation, and the triggering mechanism contain important
information on SFHs, which can be used to test EGs' formation
scenario.

The main purpose of this work is to obtain the stellar population
properties of different types of EGs (e.g., star-forming and
quiescent galaxies) and to shed light into the difference and
connection in their SFHs. We thus select a sample of local EGs with
reliable morphology classification from the Sloan Digital Sky
Survey; then different methods of stellar population analysis are
used to derive the properties of their stellar populations.

This paper is organized as following. Section 2 describes the sample
selection and data reduction in this work. In Section 3, we describe the
process of stellar population analysis and the method of emission
line diagnostic, and Section 4 presents the main results,
including the basic properties of the sample and the information
about their stellar population.  In Section 5, we discuss the
implication of the results in the context of elliptical galaxies
evolution, Section 6 is the conclusion.

Throughout this paper, we assume the $\Lambda$CDM cosmology
consistent with the \emph{Wilkinson Microwave
Anisotropy Probe}  (\emph{WMAP} ) results with $\Omega_m=0.3, \Omega_\Lambda=0.7$
and $H_0=75 h_0 km^{-1}s^{-1}$(Spergel et al. 2007) and magnitudes
are given in the AB system.

\section[]{The Sample}

The sample is extracted from Fukugita et al. (2007; F07), which
provided a catalog of morphologically classified galaxies  in a
region ($\sim$ 230 $deg^2$) of  the northern sky
 ($145\textordmasculine<\alpha<236\textordmasculine,
-1.26\textordmasculine<\delta<1.26\textordmasculine $). In this
area, there are 2658 photometric objects with r-band Petrosian
magnitude brighter than 16 mag (e.g., $r_p\leq 16$ mag) in the Sloan
Digital Sky Survey (SDSS) (Gunn et al. 1998; Blanton et al. 2003),
 among of which,  1866 have spectroscopic information.
From the spectroscopic sample, 487 galaxies which are classified as
E-E/S0 ($T=0\sim0.5$) are selected to be the sample of this work.
These 487 galaxies have both photometrical and spectroscopic
informations in the SDSS database.

In order to examine the statistical completeness, Fukugita et al.
(2007) compared the number of galaxies as a function of r-band
magnitude with the N$\sim$$10^{0.6r}$, which is expected for the
Euclidean geometry, and found that the completeness for the sample
is pretty good and does not miss too many galaxies even in the
bright end of 10 $-$10.5 mag. For the spectroscopic sample, the
completeness remains good for $r>12.5$ mag, which means we might
miss some bright galaxies. The number count of the sample EGs as a
function of r-band Petrosian magnitude, r-band absolute magnitude
and redshift are shown in Fig 1.

Among these 487 galaxies, 269 are classified as $T=0$ and the others
have $T=0.5$, whose morphological type is E/S0. After examining the
g-band images of all these 487 galaxies, it is very difficult to
tell the difference between E and S0 galaxies by eye. However, we
find that E/S0 (T=0.5) galaxies have somewhat smaller minor/major
axis ratio than EGs (T=0) galaxies, some E/S0s have a very bright
nucleus. But, in general, these two sub-samples are basically
similar in the morphology, so all 487 galaxies are taken as the EG
sample, although it may introduce contamination of S0 galaxies. The
visual classification is really important for our work, because the
correspondence between colour and morphology is complex (Lintott
et.al. 2008) and only the visual classification of elliptical
galaxies has no bias towards the ones with relative red color and
without emission line.

\begin{figure}

   \includegraphics[width=8.5cm]{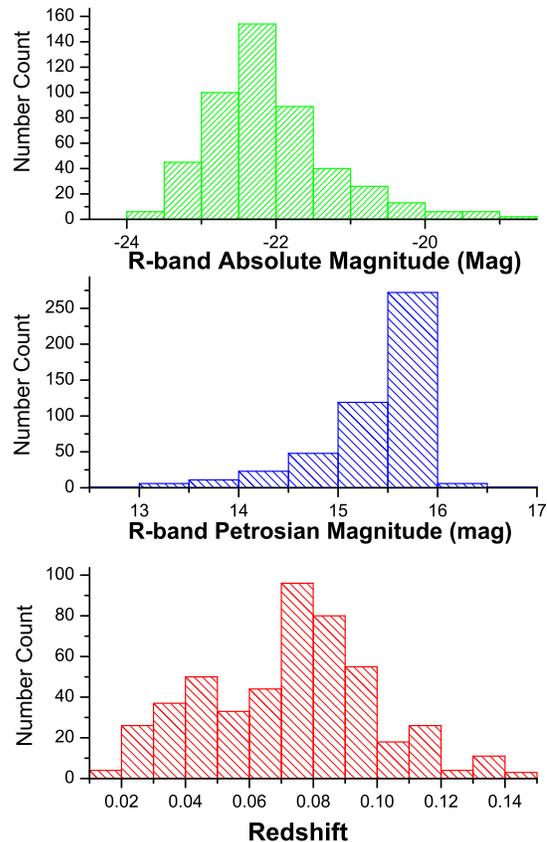}
   \caption{\textbf{The histograms of r-band Petrosian magnitude, r-band absolute magnitude and redshift of the sample}}

\end{figure}

The Petrosian magnitude, Petrosian radius (R50, R90) and other
photometric informations of five bands (\emph{u,g,r,i,z}) (Fukugita
et al. 1996; Gunn et al. 1998) are extracted from the SDSS database
\footnote{SDSS DR6: http://www.sdss.org}. The SDSS spectrum is
obtained with a 3-arcsec fiber, covers the spectral range of 3800 to
9200 \AA\ with the spectral resolution of $\sim$1800. The typical
signal-to-noise ratio (S/N) for our EGs is 40. The flux of emission
lines, 25 Lick/IDS absorption line indices and corresponding errors
are retrieved from the value-added galaxies catalog of SDSS provided
by the MPA/JHU team\footnote{MPA/JHU VAGC:
http://www.mpa-garching.mpg.de/SDSS/DR4/}. The Lick indices are
measured under the original SDSS resolution and corrected for
contamination of sky emission lines.

\section{Data Analyses}

\subsection{Stellar Population Synthesis}

In order to explore the SFH of EGs, we apply
 the stellar population synthesis code, \texttt{STARLIGHT}\footnote{\texttt{STARLIGHT} \& SEAGal:
http://www.starlight.ufsc.br/}(Cid Fernandes et al. 2005, Mateus et
al. 2006; Cid Fernandes et al. 2007; Asari et al. 2007), for EGs in
our sample. The code is based on fitting an observed spectrum
$O_\lambda$ with a linear combination of simple theoretical stellar
populations computed from evolutionary synthesis models (Cid
Fernandes et al. 2004).  The model spectrum is given by:

\begin{equation}\label{1}
   M_\lambda=M_{\lambda_0}(\sum_{j=1}^{N_\star}x_j b_{j,\lambda} r_{\lambda})\otimes
   G(v_\ast, \sigma_\ast)
\end{equation}

\noindent Where $M_\lambda$ is the model spectrum, $M_{\lambda_0}$
is the synthesis flux at the normalization wavelength, $x_j$ is the
so-called \emph{population vector}, $b_{j,\lambda}$ is the
\emph{j}th SSP spectrum at $\lambda$ and $r_\lambda \equiv
10^{-0.4(A_\lambda-A_{\lambda_0})}$ represents the reddening term.
At the end, the $G(v_\ast, \sigma_\ast)$ is the line-of-sight
stellar motions that is modelled by a Gaussian distribution centered
at velocity $v_\ast$ and with velocity dispersion $\sigma_\ast$. In
this work, the model SSPs are from the BC03 evolutionary synthesis
models(Bruzual \& Charlot 2003)which have about the same spectral
resolution of SDSS. The model base follows the work of SEAGal Group
which is made up of $N_\ast$$=150$ SSPs -- 6 metallicities (Z=0.005,
0.02, 0.2, 0.4, 1 and 2.5 $Z_\odot$)\footnote{The stellar
metallicity is defined as the fraction of mass in metal and for our
sun, $Z_\odot=0.02$} and 25 ages (from 1Myr to 18 Gyr). The spectra
were computed with Chabrier (2003) IMF, Padova 1994 models and the
STELIB library (Le Borgne et al. 2003). The intrinsic reddening is
modeled by the foreground dust model, using the extinction law of
Cardelli, Claytion \& Mathis (1989) with $R_v=3.1$.

Before fitting, the spectra are shifted to the rest frame and
corrected for the Galactic extinction according to the maps of
Schlegel, Finkbeiner \& Davis (1998) and Cardelli, Claytion \&
Mathis (1989) extinction law, then they are rebinned into 1\AA\ from
3600 to 8600 \AA\ and the spectral regions of strong emission lines
and bad pixels are masked. Fig. 2 and 3 show examples of spectral
fits for two EGs in our sample. The top panel shows the observed
spectrum (black) and the model (red). The bottom panel shows the
residual spectrum and the masked regions are plotted with the pink
line. These two examples demonstrate that the fitting method can
reproduce spectrum of EGs to an excellent degree of accuracy.

\begin{figure}
   \includegraphics[width=9cm]{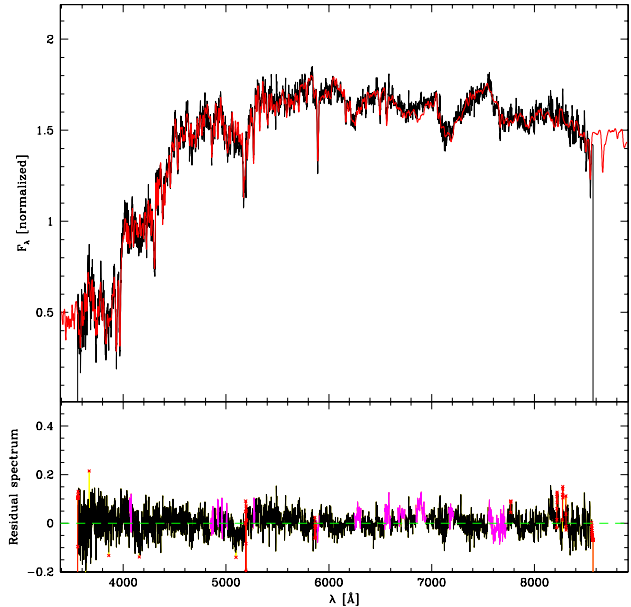}
   \caption{Spectral synthesis of an elliptical galaxy which does not show obvious emission line (SDSS J132155.18-001849.2). Top: Observed spectrum (in black line) and model spectrum
   (in red line). Bottom: The residual spectrum (black), the masked regions are plotted with a pink line. }
\end{figure}

\begin{figure}
   \includegraphics[width=9cm]{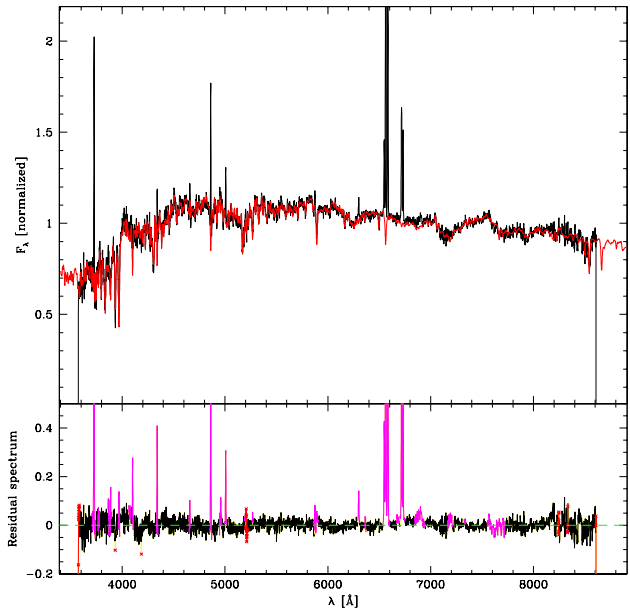}
   \caption{Same as Fig. 2, but for an elliptical galaxy which shows obvious emission lines (SDSS J101537.59+003131.0).}
\end{figure}

\texttt{STARLIGHT} presents the fraction of each stellar component,
the intrinsic extinction $A_v$, the velocity dispersion
$\sigma_\ast$, and the current stellar mass $M_\ast$. Following Cid
Fernandes et al. (2005), we could derive
 the flux- and mass-weighted average ages, which are defined as:

\begin{equation}\label{1}
   <\log t_\ast>_L=\sum_{j=1}^{N_\star}x_j \log t_j
\end{equation}

\noindent where $x_j$ is the \emph{population vector} (The fraction
of flux contributed by certain SSP) and

\begin{equation}\label{1}
   <\log t_\ast>_M=\sum_{j=1}^{N_\star}\mu_j \log t_j
\end{equation}

\noindent The $\mu_j$ is the fraction of stellar mass contributed by
each SSP. The flux- and mass-weighted average metallicities
($<Z_\ast>_L$ and $<Z_\ast>_M$) are derived in the same way.

\subsection{Emission Line Classification}

In the previous work, such as Fukugita et al. (2004); Zhao et al.
(2006) and  \textbf{Schawinski et al. (2007, hereafter S07)},
several EGs have been detected with residual star-forming
activities. In order to isolate weak AGNs, EGs are classified by
emission lines according to their ionization states estimated from
the flux ratios (\textbf{Baldwin, Philips \& Terlevich 1981};
Veilleux \& Osterbrock 1987). The emission line ratios used are
 [O III]$\lambda$5007 /\Hb,   [N II]$\lambda$6583/\Ha ,
  [SII]$\lambda$6716+6731/\Ha \ and  [O I]$\lambda$6300 /\Ha.
 Following the
S/N criterion of Kauffmann et al. (2003),  we request S/N$>3$ for
all the four lines: H$\alpha$, H$\beta$, [OIII]$\lambda$5007 and
[NII]$\lambda$6583. For [SII]$\lambda$6716+6731 and [O
I]$\lambda$6300, we did not set any criteria on their S/N since they
are basically very weak in EGs.

In the sample, 267 EGs with their S/N of H$\alpha$ emission line
greater than 3 (54.8\% of the sample), 84 EGs with S/N(H$\beta$)$>3$
(17.4\%), 238 EGs with S/N([OIII]$\lambda$5007)$>3$ (48.9\%) and 284
EGs with S/N([NII]$\lambda$6583)$>3$ (58.3\%). \textbf{The fraction
of EGs with obvious emission lines, which are defined as EGs have
S/N$>3$ for all four lines in the first BPT diagram, is $15.0\%$.
}For comparison, such fraction is about 5 $\%$ and $18\%$ for
Bernardi et al. (2003a,b) and K07, respectively. For 73 EGs with
obvious emission lines, the diagnostic diagrams are (Fig. 4) used to
classify them into different types according to the widely used
criteria of Kewley et al. (2006), where we detect 14 star-forming
EGs, 50 AGNs and 8 composite EGs. All the other EGs which are not
satisfied our S/N criteria are classified as the quiescent EGs,
although, it should be noticed that the fraction of EGs have
S/N(H$\alpha$)$>3$ (54.8\%) is much higher than the fraction of EGs
used for BPT diagrams, which means among the quiescent EGs, many of
them could have very low level of activity such as AGN or recent
star formation and may be compatible with the retired galaxy model
proposed recently by Stasinska et.al. (2008). Also, it should be
mentioned that the composite EGs are classified only from the first
emission line diagnostic diagram in Fig.4.

\begin{figure*}

   \includegraphics[width=17cm]{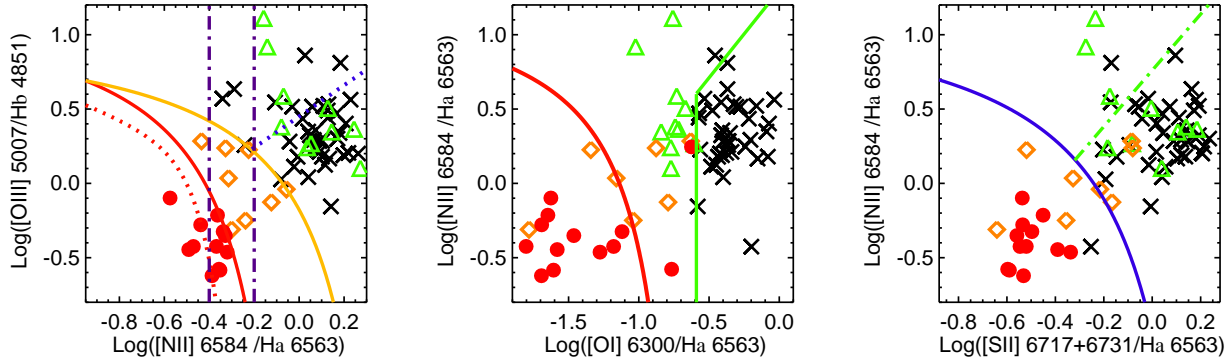}
   \caption{Emission line diagnostic diagrams from BPT. The solid lines in the diagrams are from the criteria of
Kewley et al. (2006), the red dash line in the first diagram is a
more rigorous criterion from Stasinska et al. (2006) and the purple
dashed lines are the criteria from Stasinska et al. (2006) that only
used the ratio of [NII]$\lambda$6583/H$\alpha$. In these figures,
SFEGs are red circle, transition EGs are orange diamond, Seyferts
are green triangle and LINERs are black cross}

\end{figure*}

\subsection{Absorption Line Indices}

The absorption line indices are also used here for the study of the
stellar properties of EGs. The most widely used system is the
Lick/IDS system (Worthey et al. 1994, Worthey and Ottaviani, 1997,
Trager et al. 1998), which defines 25 indices including both atomic
features (e.g H$\beta$, Ca4227, Fe5270) and molecular bands (e.g Mg,
CN, TiO).

The LICK/IDS indices and corresponding errors are retrieved from the
MPA/JHU Garching DR4 VAGC as mentioned before. The measurements are
performed under the original SDSS spectral resolution. Following the
equations from Kuntschner (2004), we corrected the effect of
velocity dispersion broadening. It's worth noting that such
measurement does not satisfy the requirement of Lick/IDS system as
mentioned by Worthey et al. (1994). So the results here will not be
compared with the previous works based on Lick/IDS system. However,
by using LICK/IDS indices, we can explore relationships between the
properties of different type of EGs and verify the validity of the
results from \texttt{STARLIGHT} fitting.

\subsection{E+A galaxies}

E+A galaxies were first discovered by Dressler \& Gunn (1983; 1992)
during the research of distant galactic clusters, which show strong
Balmer absorption lines (A-type stars) and no emission lines ([O]II
$\lambda$3727) just like the spectra of normal EGs. The strong
Balmer absorption lines reveal recent starburst which happened in
about 1 Gyr (due to the lifetime of A-type stars). So the E+A
galaxies are often considered as the post-starburst galaxies
(Kaviraj et al. 2007; Yang et al. 2008). In some work, \textbf{such
galaxies are also called as K+As (Franx 1993, Dressler et.al. 1999)
due to the presence of an old component in the galaxy spectra,
resembling a K star spectrum. }Although many K+A galaxies have
disc-like morphology, Goto et al. (2003) found that E+A galaxies
with higher completeness from SDSS DR5 generally have ETG-like
morphology. The E+A galaxies with ETG- or EG morphology could
possibly be the progenitors of normal EGs and can tell us some
undiscovered information about the evolution of EGs.

According to the recent works of large sample of E+A galaxies like
Goto et al. (2007a, 2007b, 2008) and  Helmboldt et al. (2008, K+A),
the selection criteria are: 1) the ($H_\delta$A) index has value
$>2.5$\AA, 2) EW([OII]$\lambda$3727)$<2.5$\AA, and 3)
EW(H$\alpha$)$<3.0$\AA. Generally speaking, these criteria are
looser compared with the ones by Goto (2007), howerer since the
EW(H$\alpha$) is included, there will be no contamination from
H$\alpha$ emission galaxies and the classification of E+A EGs in our
sample is reliable.

\section{Results}

\subsection{Spectral Classification}

By using the emission line ratios, we classify EGs with emission
lines into AGNs and star-forming galaxies. The BPT diagnostic
diagrams are shown  in Fig.4 and the statistical results are
summarized in Table 1. The effective classification is mostly
according to the first and the second diagrams. The solid lines
represent the criteria from Kewley et al. (2006), the red dot line
in the first diagram is a more rigorous criterion from Stasinska et
al. (2006). When we use this criterion, the number of SFEGs will
drop to 5. The two purple dashed lines in the first diagrams are the
criteria from Stasinska et al. (2006) which only use the
[NII]$\lambda$6583/H$\alpha$ ratio, from these criteria the number
of SFEGs is also 5 for the sample. So, it is possible that among the
13 SFEGs, the contamination from AGN still exists for some EGs. But,
in general, the classification is reliable, the active star
formation must take place in these SFEGs at different level.
\textbf{Also there is a blue dotted line in the first diagram
representing the empirical criterion from Kauffmann et.al. (2003).}

\begin{table}
\caption{The Classification Result from Emission Line Diagnostic
Diagrams} \label{tab:1} \centering
\begin{tabular}{lcc}
  \\
  \toprule
  Classification & Number & Fraction \\
  \midrule

  Star-Forming & 14 & 2.9\% \\
  Quiescent & 415 & 85.2\% \\
  Composite & 8 & 1.6\% \\
  LINER & 40 & 8.2\% \\
  Seyfert & 10 & 2.1\% \\
  E+A & 11 & 2.1\% \\
  \bottomrule
\end{tabular}
\end{table}

We have checked the optical images, historical literatures and
multi-wavelength data from {\it NED} \footnote{NED:
http://nedwww.ipac.caltech.edu/} and {\it Aladin} \footnote{Aladin
Java Applet: http://aladin.u-strasbg.fr/}
 for each SFEG and E+A galaxy. One galaxy
in the SFEGs sample (SDSS J114013.23-002442.2 or Mrk 1303) is picked
out because it is classified as a BCD (Blue Compact Dwarf) galaxy
from previous observation (Gondhalekar et.al. 1998). Though its
morphology is regular and elliptical like, this galaxy has unusual
blue optical color and the largest H$\alpha$ flux. Due to the
difference between EGs and BCD galaxies in the evolution
perspective, it is excluded from the SFEGs sample. For the remaining
13 SFEGs, the possibility of contamination by BCD galaxies is ruled
out from their images and historical records from the website. As
mentioned in the sample section, this EGs sample is actually made of
two sets of galaxies (T$=0$: EGs and T$=0.5$: E/S0).  7 SFEGs have
T$=0$ and the other 6 have T$=0.5$, meaning there are some
uncertainties in their morphological classification. But they are
still considered as SFEGs because their morphology of g- and r-band
images are very elliptical like, it is impossible to divide them
into EGs and S0s at all.

The fraction ratio of different types of EGs, Quiescent:
Star-Forming: Transition objects: Seyfert: LINERs are $85.2: 2.9:
1.6: 2.1: 8.2$, respectively. Compared with the distribution from
S07 (81.5: 4.3: 6.9: 1.5: 5.7), the fraction of quiescent EGs is a
little higher while the fraction of SFEGs is somewhat lower in our
sample, this probably reflects the sample difference between ours
and S07, e.g. a mostly pure EGs sample versus an ETGs sample. The
intrinsic fraction of EGs with active star formation should be
higher in S0 galaxies than EGs. For the transition objects and LINER
EGs, the fractions are also different. In S07, the authors discussed
the possibility of misclassifying some LINERs into composite EGs, so
it is reasonable to say the criterion here is favorable for the
selection of LINERs after carefully removing old stellar
contribution. After putting LINERs and transition objects together,
the fraction is 9.8\% in this work and 12.6\% in S07, the
discrepancy is much smaller. In conclusion, our results are
basically consistent with S07, our sample can be regarded as a
sub-sample of the morphogically-selected ellipticals  in S07, so the
comparability is apparent.

At the same time, we find 11 E+A EGs (fraction is $2.3\%$).
E+A galaxies are quite rare in the local universe, Goto et al.
(2003) estimated the fraction is  $\sim 1\%$. Our fraction is a
little higher since our selection criterion is
looser (EW(\Hd)$>5$\AA).

\subsection{Color - Magnitude Relations}

\begin{figure}
   \includegraphics[width=9cm]{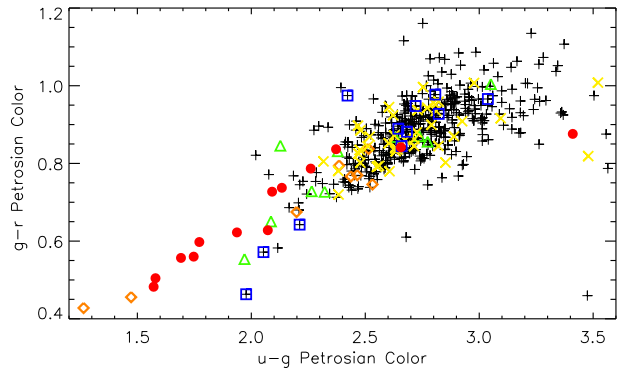}
   \caption{u-g and g-r color-color diagram. SFEGs are red circle, and E+A galxies are blue square,
The transition region galaxies are orange diamond, Seyferts are
green triangle and the quiescent EGs are black cross. These symbols
are used in the same manner from this figure }
\end{figure}

Fig.5 shows the color-color diagram of u-g vs g-r Petrosian color,
the colors have already been corrected for extinction. There are
several interesting features in this figure: 1)  The distribution of
quiescent EGs and LINERs concentrates in the same area, both of
which have red optical colors as expected for normal EGs. 2) The
distribution of SFEGs is quite scattered, $11/13$ have relative blue
colors, since they show active star-formation as found. The other
two SFEGs have optical colors just as red as the quiescent EGs, one
of which even locates among the reddest EGs. 3) For the E+A EGs, 8
E+A EGs have color distribution in the same area with quiescent EGs,
while the other 3 are quite blue.

\begin{figure}
  \includegraphics[width=9cm]{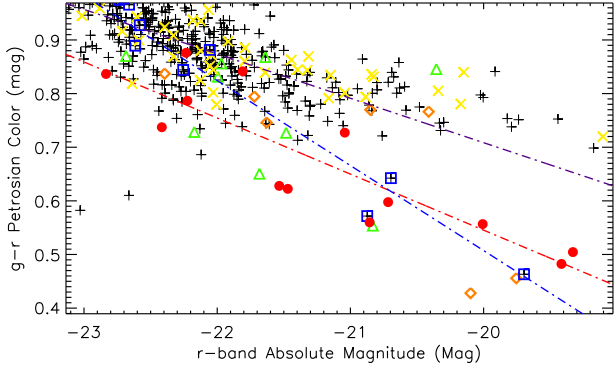}
   \caption{Color-Magnitude relation using the g-r color. The dot-dashed lines are the simple linear fitting
of the relation for the quiescent, E+A and star-forming EGs,
respectively }
\end{figure}

\begin{figure}
  \includegraphics[width=9cm]{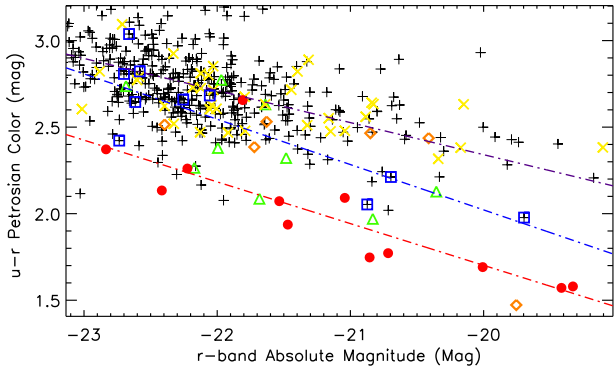}
   \caption{Color-Magnitude relation using the u-r color. The dot-dashed lines are the simple linear fitting
of the relation for the quiescent, E+A and star-forming EGs,
respectively }
\end{figure}

The color-magnitude relation (CMR)  is a famous scaling relation for
EGs (Bower, Kodama \& Terlevich 1998), the tightness is considered
as evidence for the \emph{monolithic collapse} scenario. Fig.6 is a
plot of CMR, where the x-axis is the r-band absolute magnitude which
has been extinction and k-corrected, and the y-axis is the g-r
Petrosian color. Three dot-dashed lines are the simple linear
fitting of the relation for the quiescent, E+A and SFEGs,
respectively. The relation is basically tight for the whole sample,
while the scatter is quite large
 at the low-luminosity end. The
slopes of the CMR for SFEGs and quiescent EGs are similar but the
zero point are different significantly, meaning that the SFEGs tend
to have bluer colors than the quiescent EGs with the similar $M_r$.
The SFEGs have much lower average $M_r$ than the quiescent EGs, 5 of
which are fainter than $M_r=-21$ mag. In Fig.7, we plot r-band
absolute magnitude versus the u-r Petrosian color, the trend is the
same but the scatter is even larger than Fig.6. The relations of
SFEGs and quiescent galaxies are well separated. The CMR used to be
interpreted as a sequence of increasing metallicity with luminosity.
Now we already know, in addition to metallicity, the stellar age
also plays an important role. Based on a very large sample of SDSS
ETGs, Gallazzi et al. (2006) found unambiguous evidence for
interpreting the CMR as a sequence of galaxy stellar mass.

\begin{figure*}
  \includegraphics[width=15cm]{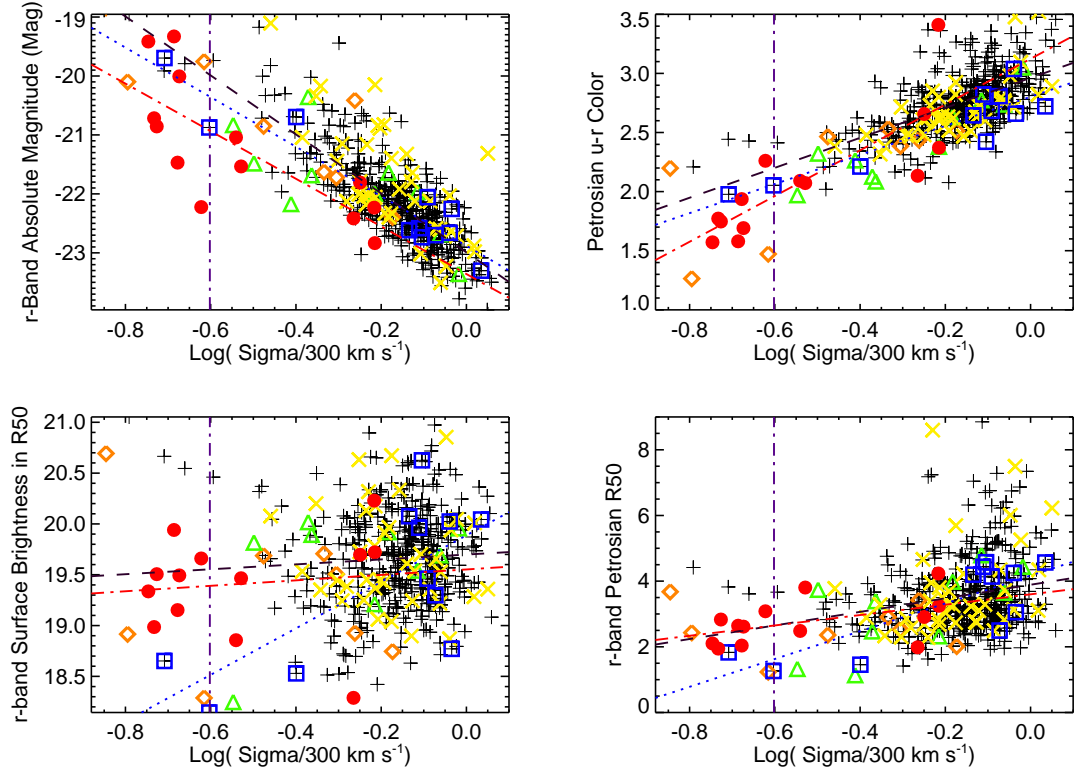}
   \caption{The relation between the velocity dispersion and the galaxy absolute magnitude, u-r color,
r-band surface brightness and the r-band Petrosian R50. The lines
are the linear fitting of these relation for the quiescent,
star-forming and E+A galaxies. The black dash line is for the
quiescent EGs, the red dash dot line is for the SFEGs and the blue
dot line is for the E+A EGs. The vertical dash dot line represent
the SDSS spectra dispersion (75 km/s).}
\end{figure*}

\begin{table*}
\caption{The average general properties for different types of
elliptical galaxies. \textbf{The values in parenthesis are the
corresponding standard deviation}.} \label{tab:2} \centering
\begin{tabular}{cccccc}
       \\
       \toprule
       Properties\textbackslash Classification & Quiescent & LINER &
       Seyfert & E+A & Star-Forming \\
       \midrule
       r (mag) & 15.01(0.58) & 15.29(0.60) & 15.49(0.54) & 15.42(0.48) & 15.58(0.35)\\
       M (Mag) & -22.17(0.76) & -21.80(0.92) & -21.82(0.85) & -22.02(1.11) & -21.06(1.23)\\
       u-r (Mag) & 2.75 (0.30)  &  2.70 (0.25) & 2.43 (0.35) &
2.55 (0.34) & 2.10 (0.51) \\
       SB ($Mag/(arc sec)^2$) & 19.66(0.48)& 19.68(0.49) & 19.40(0.80) & 19.42(0.80) & 19.40(0.48) \\
       CI & 3.17(0.22) & 3.21(0.19) & 3.00(0.32) & 3.14(0.29) & 2.89(0.33)\\
       R50 (arc sec) & 3.55(1.23) & 3.71(1.41) & 3.13(1.26) & 3.30(1.32) & 2.71(0.71)\\
       \bottomrule
\end{tabular}
\end{table*}

\begin{figure}
  \includegraphics[width=9cm]{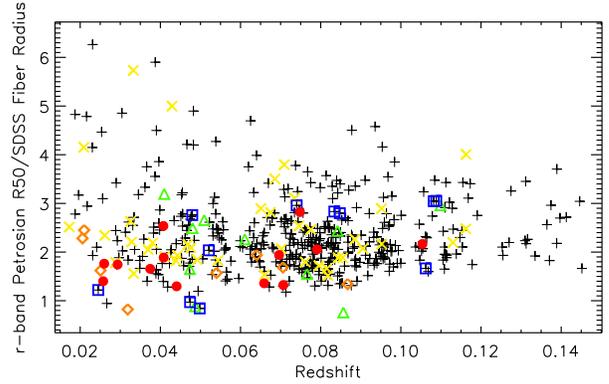}
   \caption{The ratio of Petrosian R50 and SDSS fibre size as a function of the redshift.}
\end{figure}

For EGs or ETGs, the central velocity dispersion is a wildly
accepted indicator for galaxy mass. The \texttt{STARLIGHT}  can
provide us  the velocity dispersion (Cid Fernandes et al. 2006).  We
do not set a lower limit for the velocity dispersion during the
\texttt{STARLIGHT} fitting, while the SDSS spectra have intrinsic
instrumental dispersion of ~75 km/s (York et.al. 2000). Some EGs
show velocity dispersion lower than this value, this of course will
lead a serious uncertainty. For these EGs, their velocity
dispersions are still used. Since the velocity dispersion is treated
as a Guassian with broad wings larger than the instrumental
dispersion, thus the code can measure velocity dispersion slightly
below 75 km/s (Schawinski, private communication). But we will keep
this uncertainty in mind and just take these measurements as very
rough estimation. And the velocity dispersion is corrected to the
r-band half-light radius (R50) according to the equation in
Cappellari et.al.(2006).

Fig.8 show the relations between the velocity dispersion
($\sigma_\ast$) and several fundamental properties of the EGs: the
r-band absolute magnitude, the u-r Petrosian color, the r-band
surface brightness and the r-band Petrosian R50. 16 EGs have
$\sigma_\ast <$ 75 km/s: 7 SFEGs, 2 E+A EGs. The vertical dash-dot
line represents the intrinsic dispersion of SDSS spectrum (e.g.
$\sim$ 75 km/s). More than 50\% SFEGs show $\sigma_\ast$ less than
75 km/s, which make further analysis more unreliable, though it
might suggest that the SFEGs have smaller stellar mass in the local
universe.

From these figures, we see tight relations of $\sigma_\ast$-$M_r$,
$\sigma_\ast$-$(u-r)$, which could be explained as more massive EGs
are more luminous and redder. The scatter of the $\sigma_\ast$-$M_r$
relation is larger below the 75km/s boundary. For the
$\sigma_\ast$-$u-r$ relation, the scatter keeps about the same along
the relation. The slopes of the correlations are almost same for all
types of EGs, suggesting that these correlations may involve some of
the most fundamental properties of the EGs and are barely influenced
by their activity. Generally speaking, the SFEGs tend to be less
massive, bluer and fainter than the normal quiescent EGs, which is
consistent with S07. No correlation was found between the
$\sigma_\ast$ and the surface brightness. It is physical reasonable
for more massive EGs have bigger size, but the Petrosian R50 should
be taken as the proxy of the effective radius of EGs, not exactly
physical size. No clear correlation is found between the galaxy mass
and their effective size, which could be explained by different
brightness profile along the radius, which could be the evidence of
different SFH for EGs have different mass.

The average properties and scatters of basic properties for
different types of EGs are listed in Table.2, including the r-band
Petrosian magnitude, r-band absolute magnitude, the u-r Petrosian
color, the r-band surface brightness, concentration index and
Petrosian R50. Except for the quiescent EGs, the sample size for
other types of EGs is quite small, thus the statistical results are
meanless. On average, the SFEGs have the faintest $M_r$, the bluest
$u-r$ color, the smallest Petrosian R50 and concentration index.
From S07, the authors discussed the possibility of an evolutionary
sequence from star-forming EGs to the quiescent EGs. Considered the
"Down-Sizing" effect which means more massive EGs should evolve
faster than their less massive counterparts, the trend is also seen
here.

Now it is necessary to mention the problem
of aperture effect of fixed SDSS 3-arcsec fiber, which
means the physical information extracted could be
only accounted for the central part of the EGs.
 Fig.9 shows a plot of the ratio of Petrosian R50
and SDSS fibre size as a function of the redshift. Although some EGs
 do suffer from aperture effect, in
general, there is no strong aperture bias for the whole sample,
especially for SFEGs. Most EGs have r-band Petrosian R50 1-3 times
larger than the SDSS fibre radius, but it still worth noting that
the physical information about the stellar population obtained below
should only accounted for the region within $1/3$ to 1 effective
radius of the EGs.

\subsection{Star-Forming Elliptical Galaxies}

\begin{figure}
\begin{center}
\includegraphics[width=7cm]{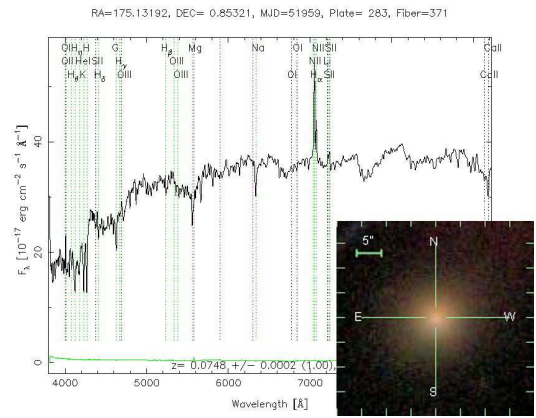}
\caption{SDSS J103534.47-002116.2}
\end{center}
\end{figure}

\textbf{1. SDSS J114031.65+005111.4  (PGC1177304)}

The morphology of this galaxy is similar to a typical elliptical
galaxy, the optical color is also close to the quiescent EGs
(g-r$=0.79$), but strong emission lines are detected in the SDSS
spectrum. This galaxy is $305.6\pm21.4$ Mpc away, the r-band
absolute magnitude $M_r=-21.40$. Based on Tago et al. (2008), it
belongs to a small galaxy group (Group944) which has 4 members. This
galaxy has 2MASS (Skrutskie et al. 2006) (J=14.826, H=14.112,
K=13.596) and GALEX (Martin et al. 2005) (FUV=20.14) observation.
The image and spectrum are shown in Fig.10.

\begin{figure}
\begin{center}
\includegraphics[width=7cm]{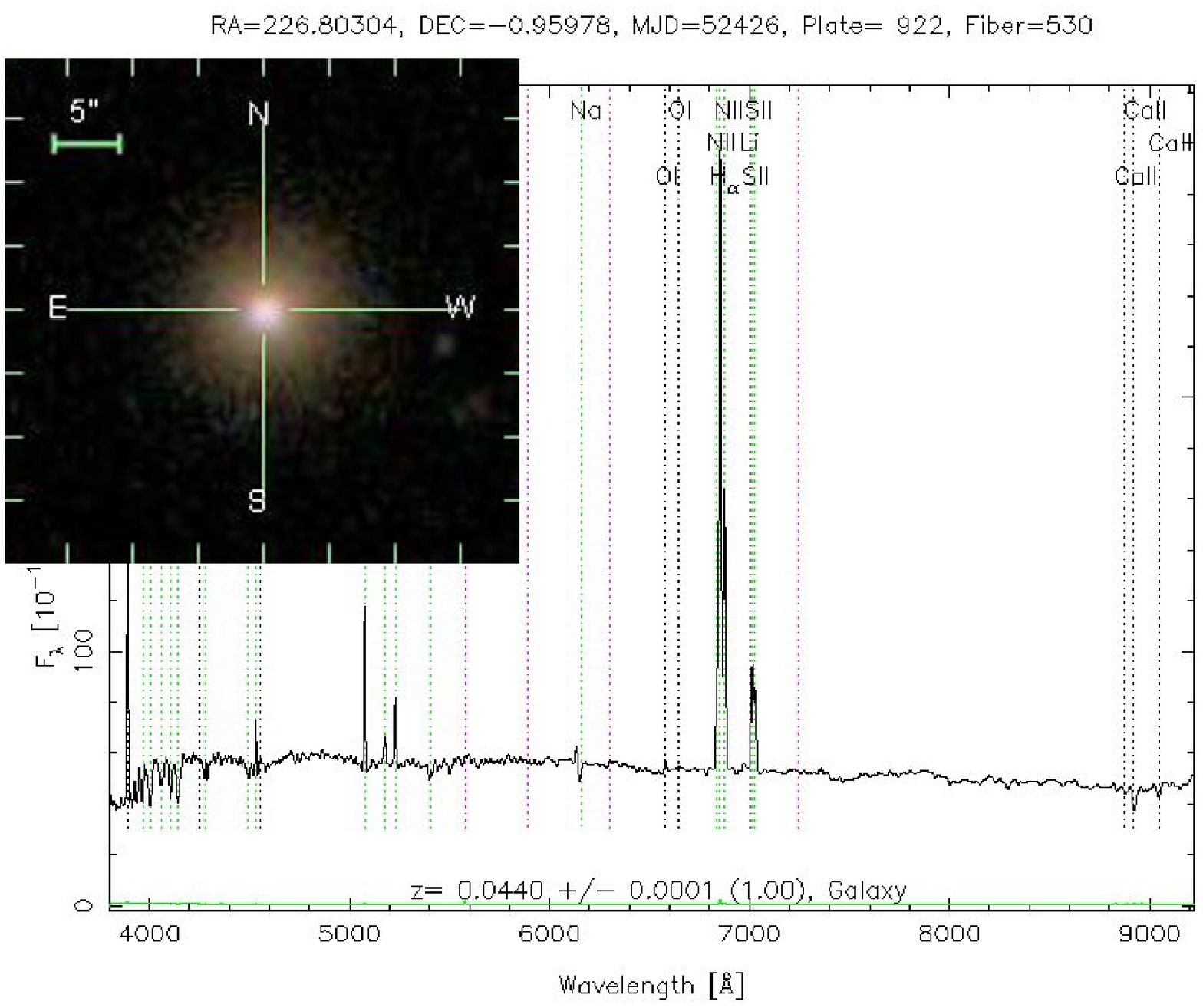}
\caption{SDSS J150712.72-005735.2}
\end{center}
\end{figure}

\textbf{2. SDSS J150712.72-005735.2 (PGC1130399)}

It is also a typical elliptical galaxy in morphology, but has very
blue (g-r=0.50) optical color and a very bright nucleus. The
spectrum is very similar to star-forming galaxies with flat continua
and strong emission lines. This galaxy is classified as emission
line galaxy (ELG) in NED. It is $181.2\pm12.7$ Mpc away and bright
absolute magnitude ($M_r$=-23.05). This galaxy does not belong to
any galaxy group or cluster. The 2MASS J, H and K magnitudes are
14.72, 14.11, and 13.56 mag, respectively. Its image and spectrum
are given in Fig.11.

These two galaxies represent the SFEGs sample well and the detailed
information for the other 11 SFEGs are listed in Table 3. Since this
sample only belongs to a small sky region, it is not good for
environmental research. We find that none SFEGs is cluster member, 3
SFEGs belong to small galaxy group,  and 2 are even classified as
isolated galaxies by Allam et al. (2005).

\begin{table*}
\caption{The Properties of the 13 Star-Forming Elliptical Galaxies.}
\label{tab:3} \centering
\begin{tabular}{cccccccccc}
       \\
       \toprule
       Galaxy &  z &  $m_r$ & $M_r$ & g-r & $SB_{R50}$ & $R_{Petro}$ &  $\log(O/H)+12$ & $Av_{Nebular}$ &
       SFR \\
       \midrule
       SDSS J120823.51+000636.9 & 0.041  & 14.783  & -21.534  & 0.697 & 19.465  & 8.843  & 9.135  & 1.016  & 0.332\\
       SDSS J103534.47-002116.2 & 0.029  & 15.568  & -20.007  & 0.464 & 19.494  & 6.529  & 9.016  & 0.575  & 0.295 \\
       SDSS J152347.10-003823.0 & 0.037  & 15.089  & -21.042  & 0.628 & 18.856  & 6.226  & 9.180  & 1.142  & 0.321 \\
       SDSS J123807.26+011446.1 & 0.026  & 15.862  & -19.416  & 0.503 & 19.336  & 4.614  & 9.033  & 0.317  & 0.218 \\
       SDSS J130029.44+003556.0 & 0.079  & 15.645  & -22.224  & 0.685 & 19.660  & 6.572  & 9.160  & 1.021  & 0.299 \\
       SDSS J114031.65+005111.4 & 0.075  & 15.497  & -22.234  & 0.792 & 20.229  & 9.875  & 9.148  & 1.348  & 0.239 \\
       SDSS J140622.76-001325.1 & 0.105  & 15.720  & -22.835  & 0.701 & 19.721  & 8.340  & 9.027  & 1.250  & 0.678 \\
       SDSS J112326.98-004248.8 & 0.041  & 15.467  & -20.856  & 0.519 & 19.506  & 6.434  & 9.064  & 1.055  & 0.550 \\
       SDSS J112531.57+002619.1 & 0.026  & 15.971  & -19.329  & 0.437 & 19.941  & 6.066  & 8.946  & 0.918  & 0.286 \\
       SDSS J100920.32-001854.1 & 0.070  & 15.749  & -21.808  & 0.791 & 19.697  & 6.550  & 8.685  & -3.280 & 0.014 \\
       SDSS J110741.94+002608.3 & 0.066  & 15.958  & -21.470  & 0.594 & 19.152  & 4.193  & 9.104  & 1.772  & 1.624 \\
       SDSS J150712.72-005735.2 & 0.044  & 15.777  & -20.717  & 0.500 & 18.985  & 4.933  & 9.076  & 1.066  & 0.957 \\
       SDSS J101537.59+003131.0 & 0.071  & 15.189  & -22.416  & 0.652 & 18.289  & 4.183  & 9.062  & 0.891  &
       1.478 \\
       \bottomrule
\end{tabular}
\end{table*}

In Table.3, we summarized physical properties of 13 SFEGs. Besides
the properties from SDSS data, the star-formation rate (SFR) and the
nebular metallicity are also estimated. For calculating the SFR, the
flux ratio of H$_\alpha$ and H$_\beta$ ($F_{H_\alpha}/F_{H_\beta}$)
is used to estimate the nebular extinction $A_{V,nebular}$ which is:

\begin{equation}
   A_{V,nebular}=6.31*\log(F_{H_\alpha}/F_{H_\beta})-2.88
\end{equation}

\noindent The mean nebular extinction is $A_{V, nebular}=0.70$ (and
the median is 1.02). After correcting extinction, H$_\alpha$ flux is
used to calculate the H$_\alpha$ luminosity. The SFR is estimated
following the equation from Mateus et al. (2007), which considered
the aperture correction for SDSS 3 arcsec fibre spectra:
\begin{equation}
   SFR(H_\alpha)(M_\odot/yr)=5.22\times10^{-42}L(H_\alpha)10^{-0.4(r_{Petro}-r_{fiber})}
\end{equation}

\noindent The $r_{Petro}$ is the r-band Petrosian magnitude and the
$r_{fiber} $ is the r-band fiber magnitude. The average SFR is
0.56$M_\odot/yr$ and the highest SFR$=1.62M_\odot/yr$, which are
comparable with the normal star-forming galaxies. This is absolutely
strange from the traditional view of EGs, though the SFR estimated
here is quite cursory.

Following Denicolo et al. (2002), we estimate the nebular metallicity by using
the so called "N2 ratio calibrator" ($\log([NII]6584\AA)/H_\alpha$), which is :

\begin{equation}
   12+\log(O/H)=9.23(\pm0.02)+0.79(\pm0.03)\times N2
\end{equation}

The average nebular metallicity is $\log(O/H)+12=9.05$. Among 13
SFEGs, J100920.32-001854.1 shows the lowest SFR (only 0.014
$M_\odot/yr$) and nebular metallicity ($\log(O/H)+12=8.68$, even
lower than Solar metallicity) and a negative nebular extinction (the
only SFEG with negative extinction). This galaxy also has very red
optical color and its spectrum is similar to a typical elliptical
galaxy spectra, the emission lines are unconspicuous in the observed
spectrum, it belongs to a group with 18 members (Group 265 in Tago
et al. 2008s).

\subsection{SFH from \texttt{STARLIGHT} Fitting}

By using \texttt{STARLIGHT} to fit the observed spectra of EGs, we can derive :
 1) Velocity dispersion; 2) Extinction;
3) Average stellar population age; 4) Average stellar population
metallicity and 5) the fraction of flux contribution for each SSP
(the population vector). The average stellar population age and
metallicity are estimated by both flux-weighted and mass-weighted.
These properties can be used to draw a useful picture for SFH
research. Of course, the situation here is not so optimistic, there
are still several problems. Leaving the insufficiency of the models
aside, the method of population synthesis used here has some
drawbacks itself. Besides the velocity dispersion estimate which
under the SDSS resolution, in the fitting process of
\texttt{STARLIGHT}, the code uses a universal extinction law for all
SSPs; Some of the EGs, especially for the quiescent EGs, have
negative extinction. This is all due to the problems within STELIB,
which makes the code believes the observed spectra are too "blue".
So, the value of the extinction will not be taken seriously but its
relation with other stellar population properties is still useful.
The age-metallicity degeneracy will certainly influence the
estimation of stellar population age and metallicity, it is not
realistic to derive an accurate SFH, the reliable parameters include
mean age and metallicity. During fitting, there is no detailed
restriction for different SSPs, the flux contribution from each SSP
is relatively arbitrary. The old-age high-metallicity SSPs could
contribute a lot in the total flux, which is very suspicious from
the chemical evolutionary theory. As stated above, these problems
will make the results with larger uncertainties, however there are
still lots of useful information which can be extracted from the
fitting.

\begin{figure}
 \includegraphics[width=9.5cm]{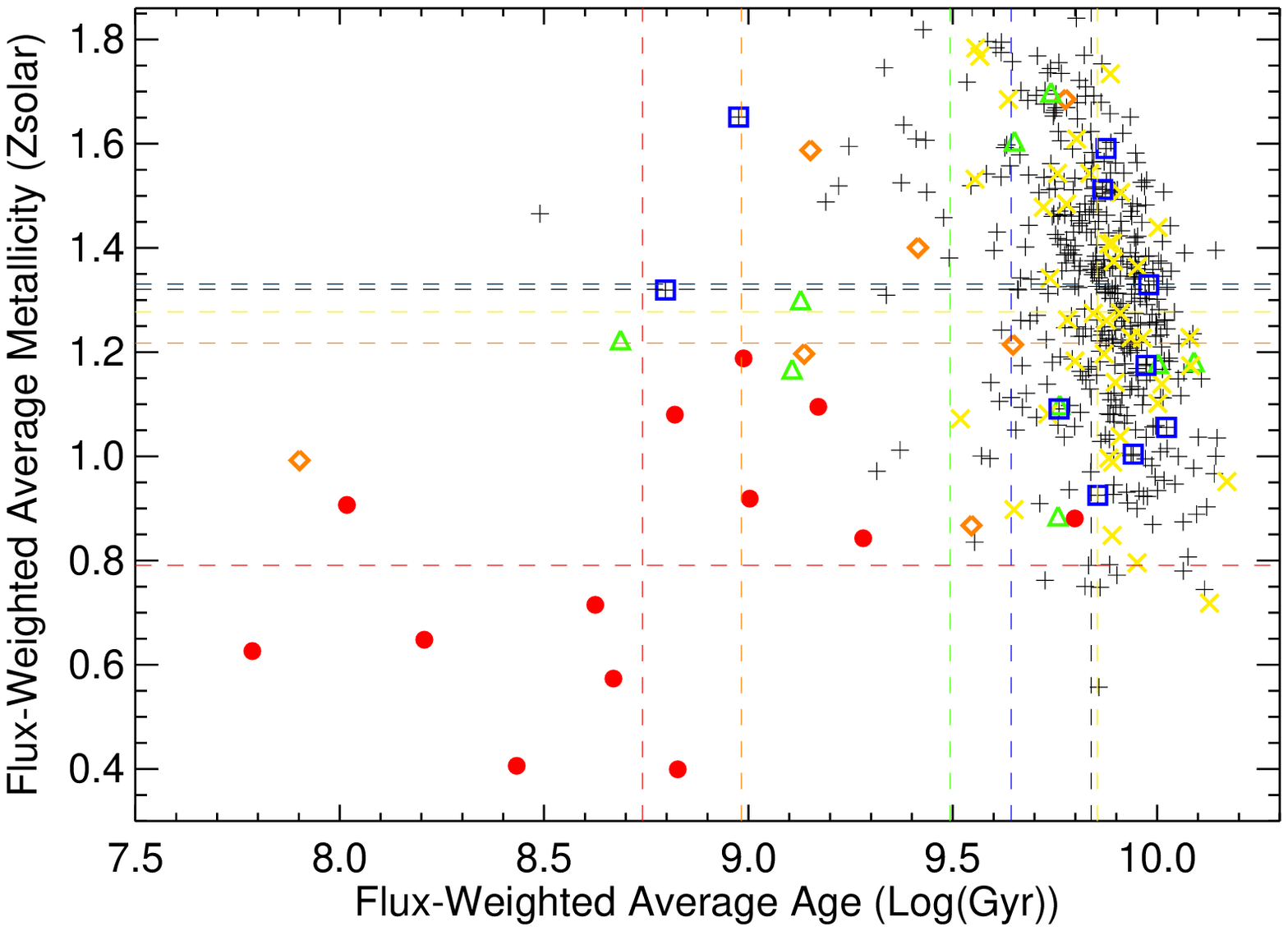}
   \caption{The distribution of the average age and metallicity for different type of elliptical galaxies. The age and metallicty
here are both flux-weighted. The horizontal and vertical dash lines
represent the average age and metallicity for different types of
elliptical galaxies.}
\end{figure}

\begin{figure}
   \includegraphics[width=9.5cm]{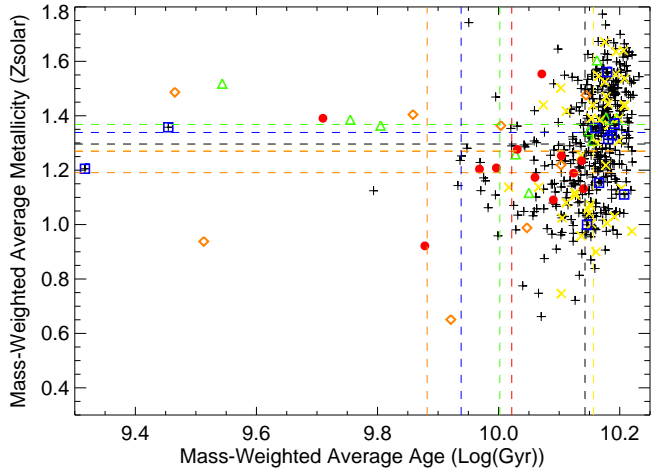}
   \caption{The distribution of the average age and metallicity for different type of elliptical galaxies. The age and metallicty
here are both mass-weighted. The horizontal and vertical dash lines
represent the average age and metallicity for different types of
elliptical galaxies.}
\end{figure}

\subsubsection{The Distribution of Mean Age and Metallicity}

In the study of stellar population in EGs, the age-metallicity
degeneracy is still an annoying problem that can not be removed
completely. The average age and metallicity from different SSPs are
more reliable statistically. The two different weighted methods
(flux- and mass-weighted) have different meaning and are related to
different properties of SFH. The average flux-weighted stellar age
is more sensitive to the younger stellar population and the residual
star formation activity. The situation is same for the flux-weighted
average metallicity. But the mass-weighted average metallicity is
more physical for EGs. In Fig.12 and 13, we plot the distribution of
these average properties in the age-metallicity plane. Fig.12 is for
flux-weighted properties and Fig.13 is for the mass-weighted ones.
The dashed lines represent the mean stellar population age and
metallicity for different types of EGs.

From Fig.12, we find the distribution of EGs is quite scatter, which
is mainly caused by the SFEGs. The distribution for other EGs are
located in a much smaller region. The SFEGs have the youngest mean
flux-weighted average age and the lowest mean flux-weighted average
metallicity, 10 SFEGs have $<Z_{flux}> \le Z_{\odot}$. For E+A EGs,
two of which show relative low flux-weighted average age and the
others are just like the quiescent EGs. We note that  E+A EGs show
the highest mean flux-weighted metallicity.

In Fig.13, the overall distribution of EGs is much more
concentrated. The distribution of SFEGs is much closer to quiescent
EGs when using the mass-weighted properties. As mentioned above, the
flux-weighted properties are more sensitive for young
 stellar population, which is always associated with recent
star-forming activity. However, SFEGs show much older mean age when
using mass-weighted population vector. Surprisingly, two E+A EGs
with lower average age around $10^9$ Gyrs in flux-weighted plot show
the lowest average age in the mass-weighted plot, even much lower
than the SFEGs. The detailed results of the synthesis indicate most
of their stellar mass are formed around $10^9$ Gyrs and almost no
old stellar population, this is absolutely strange for EGs and we do
not know for sure whether it is real or spurious.

\begin{figure*}
  \includegraphics[width=15cm]{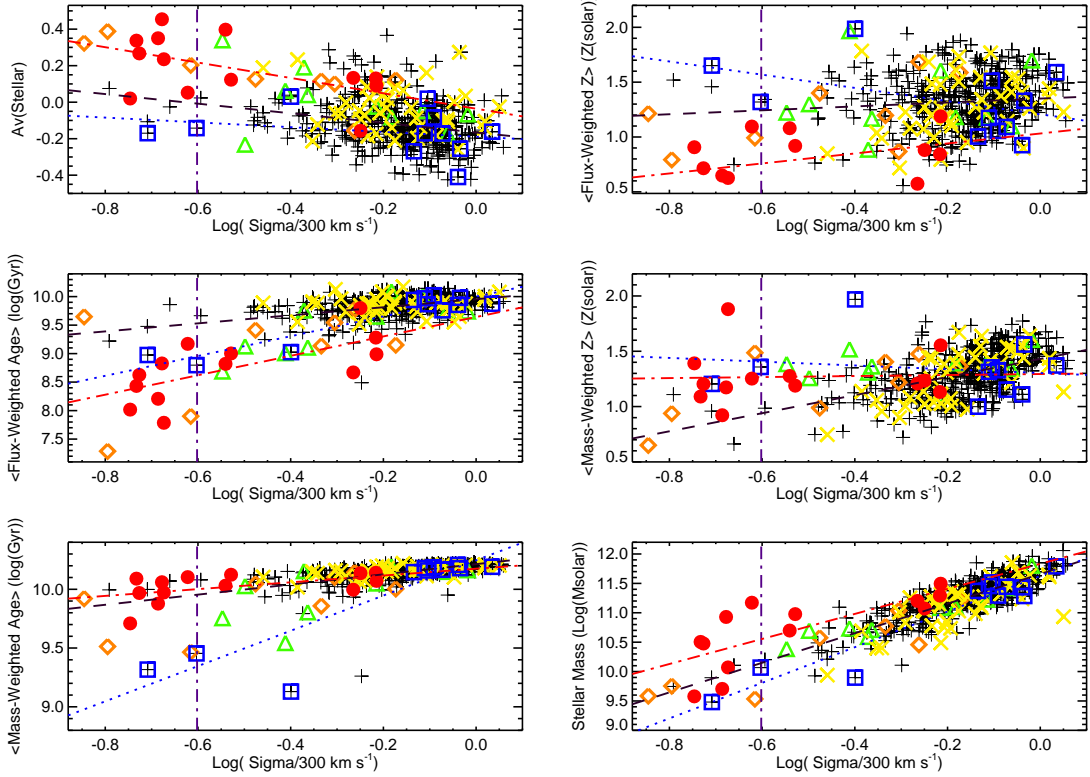}
   \caption{The correlation of different stellar population properties with the velocity dispersion for different types of EGs.}
\end{figure*}

\subsubsection{Stellar Mass}

Many works on stellar populations in EGs show that many properties
correlate with stellar mass (Thomas et al. 2005; Denicolo et al.
2005; Schawinski et al. 2007), which indicates that for EGs with
different mass, their SFH and the chemical evolution history could
be different. This is an important result for the theory of
formation and evolution of EGs . Here, we show the correlation of
the stellar population properties of EGs with their mass. The
velocity dispersion from \texttt{STARLIGHT} fitting is used as the
proxy of mass as before, so the problem with velocity dispersion
lower than SDSS spectral resolution still remains, but it will not
influence the general trend. Fig.14 shows the relations between the
$\sigma_\ast$ and some stellar populations parameters: the dust
extinction, the two groups of average age and metallicity and the
current stellar mass. We also show the linear fitting of the
relations for quiescent, E+A EGs and SFEGs, the vertical dot dash
line is still the 75 km/s limit.

The relation between the stellar velocity dispersion and the current
stellar mass is shown in Fig.14. We find that the relation is very
tight for the galaxies with $\sigma_\ast$ larger than 75 km/s, but
for EGs with $\sigma_\ast$ less than 75 km/s, the scatter is much
larger and several SFEGs are clearly away from the relation. This
uncertainty has already been predicted before and, at the same time,
we noticed that the stellar mass from \texttt{STARLIGHT} fitting
have much larger values for these EGs. Besides the uncertainty from
SSP fitting, there might be other factors. For example, from  more
detailed observations like the \texttt{SAURON} project (Bacon et al.
2001; de Zeeuw et al. 2002), there are accumulated evidence for the
existence of stellar disk in the central region for some EGs,
especially for the SFEGs (McDermid et al. 2006). This
rotation-supported disc component could lower the measurement of the
velocity dispersion. This could be just the case for the SFEGs with
very low velocity dispersion in our sample. Nevertheless, the
velocity dispersion is still used as the proxy of galaxy mass
throughout this paper.

We also find the correlation between dust extinction and
the average age, both flux- and mass-weighted. In the first plot, a
large fraction of EGs have negative $A_{V,stellar}$, especially for
the quiescent EGs. However, most SFEGs ($12/13$) show
positive $A_{V,nebular}$.
We can see a tendency that more massive EGs may have lower
dust extinction.

The correlation between stellar mass and the average stellar population age
is obvious for both flux- and mass-weighted age. This relation
indicates that the less massive EGs tend to have lower average age for
their stellar population. The relation for mass-weighted average
age is very tight except for two
special E+A EGs. We note that the mass-age relation has been reported in a lot
of works before, both in the field and the clusters (Caldwell, Rose
\& Concannon 2003; Denicolo et al. 2005; Kuntschner et al. 2002).

For metallicity, no relation is found for neither the flux- nor
mass-weighted metallicity. Maybe there is very weak trend that less
massive EGs also have lower metallicity, but the uncertainty is too
large for conclusion. However, Kuntschner et al. (2000) reported a
mass-metallicity relation for ETGs in cluster or high environmental
density region (see also Trager, Faber \& Dressler 2008).

\subsubsection{The Approximate Star Formation History}

\texttt{STARLIGHT} provides us the stellar population vector, e.g.
the fraction of flux contributed by certain SSPs. Due to the
drawbacks we discussed above, this vector does not stand for the
accurate star-formation history. In fact, our SSPs are generally
uniformly separated in the log-space of age, so the model's
resolution for stellar population analysis is limited in the end of
old age. Even more SSPs are used, the age-metallicity degeneracy and
the arbitrariness of the fitting will prevent us from deriving a
unique answer.

\begin{table*}
\caption{The average fraction of flux and mass contributed by
stellar populations of different age. The values in parenthesis are
the corresponding standard deviation.} \label{tab:4} \centering
\begin{tabular}{ccccccc}
       \\
       \toprule
       Properties\textbackslash Classification & Quiescent & LINERs & Seyfert & E+A & Star-Forming \\
       \midrule
       $f_{burst}$  & 2.72(3.25)   & 3.02(4.12)   & 7.87(9.02)   & 3.80(4.88)   & 35.79(16.92)\\
       $f_{young}$  & 5.23(8.44)   & 4.33(5.03)   & 21.67(25.32) & 19.45(26.79) & 51.57(21.17)\\
       $f_{middle}$ & 23.70(14.02) & 20.90(13.07) & 28.31(18.02) & 22.07(12.97) & 18.85(8.31)\\
       $f_{old}$    & 71.00(15.80) & 74.71(12.90) & 49.89(31.59) & 58.38(36.15) & 29.45(18.51)\\
       \midrule
       $m_{burst}$  & 0.01(0.17)  & 0.00(0.00)   & 0.04(0.11)   & 0.00(0.00)  &  0.45(0.91)\\
       $m_{young}$  & 0.43(3.48)  & 0.01(0.05)   & 5.11(10.79)  & 9.13(16.92) & 3.85(4.44) \\
       $m_{middle}$ & 5.87(8.05)  & 3.79(3.46)   & 18.26(20.02) & 16.12(24.11) & 14.54(12.29) \\
       $m_{old}$    &93.62(10.28) &96.13(3.49)   & 76.38(26.27) & 74.66(39.05) & 81.10(14.23) \\
       \bottomrule
\end{tabular}
\end{table*}

In order to estimate the SFHs of EGs,  we combine 150 SSPs into four
ages, e.g., young-, middle- and old-age stellar population plus a
population named burst-population. The young-age stellar population
includes the SSPs with age less than 1 Gyr, the old-age population
are  SSPs with age larger than 10 Gyr, and  the middle-age
population is the SSPs between them. For the burst population, we
defined its age less than 0.1 Gyr, which is extremely young for EGs
and is always connected with the recent star formation activity.
With the fractions of these four stellar populations, a very coarse
SFH can be generated and the results are shown in Table 4.

Just as expected, we can see large diversity for different types of
EGs. The quiescent and LINERs EGs are comprised by the old stellar
population, only about 5\% of their flux is from the young stellar
population. We noticed the fraction of burst-population is not zero
for neither of them (about 3\% on average), this fraction is just
similar as the resolution of the \texttt{STARLIGHT} code ($\sim$ 5
\%, Cid Fernandes et.al. 2005), thus there is no significant
evidence for presence of young stellar population in the center of
the normal or LINERs EGs (30\%-50\% of the Petrosian R50), but see
Graves et.al. (2007), where they found the age of LINERs to be
systematically younger than their quiescent counterparts, this could
be the result from different method for deciding the average age or
the different classification of LINERs. The most impressive result
here is that SFEGs are significantly different from normal EGs, more
than 50 \% flux are from the young stellar population, especially
from the recent burst-population (younger than 0.1 Gyr). We also
notice that the fraction of young-population in E+A EGs is much
higher than the quiescent and LINERs EGs, but the fraction of
burst-population is much lower than the SFEGs, this could be treated
as evidence of their post-starburst signatures. Besides these, about
20\% flux for all types of EGs is from the middle-age populations
(between 1Gyr and 10 Gyr), this could be partly from the uncertainty
of the model and the arbitrariness of our fitting. It may also
indicate that the detailed SFH of all kinds of EGs are more
complicated than what we see here and the star-formation time scale
in most EGs can extend to the middle age bin, but within this
fitting, no further information could be obtained.

We must emphasize these results are from the flux-weighted average
population fraction, which is more sensitive to the young stellar
population and star formation activity. Another important
examination is for the mass of stellar population currently in the
EGs or the so-called mass-weighted average population fraction. The
same criteria of age bins are used and the results are also
presented in the Table 4. From this part, the diversity seen from
the flux-weighted fractions becomes very weak, most of the stellar
mass is contributed by the old population (generally, more than
80\%), even for the SFEGs. The mass fraction of burst-population is
negligible in all types of EGs (less than 1\%) and the mass-fraction
of the young-population is also negligible for both quiescent and
LINERs EGs. For the SFEGs, the fractions of stellar mass in
different age bins change a lot from the flux-weighted ones, the
fraction of stellar mass formed in recently ($<$ 0.1 Gyr) is less
than 1\% although it can be measurable when compared with the other
four types of EGs and the fraction of stellar mass formed in less
than 1 Gyr is just 4\%, even less than Seyfert and E+A EGs. The
reason may be due to the small size of the sub-samples, which makes
the statistic results less reliable, but it indeed suggusts that
these SFEGs are not totally different with the normal EGs, most
stars are formed at very early age, while a small fraction of stars
formed in the residual star-formation in the central part of the
galaxy change the flux-weighted stellar properties greatly. This
phenomenon is some kind of "re-juvenile"  phenomenon which is
related with many processes that could affect the evolution EGs. For
example, the trigger of this "re-juvenile" phenomenon could be the
minor mergers (Michard 2006): the intensity and duration could be
related to the mass of the galaxy and the feedback of their central
SMBH. Also, from the quiescent and LINERs EGs to the SFEGs, the SFH
is basically getting more and more extended though this conclusion
is derived from our small sample of EGs. At the end, we should
notice the mass-weighted population fractions of E+A EGs--without
burst-population but has about 9\% of their stellar mass formed in
recent 1 Gyr, which is the highest for these 5 different types of
EGs. This result confirms the post-starburst nature of these EGs
again and their SFH could be quite different compared with the
quiescent EGs although they have many similarities.

\begin{table*}
\caption{The Average Stellar Population Properties for Different
Types of Elliptical Galaxies. The values in parenthesis are the
corresponding standard deviation.} \label{tab:5} \centering
\begin{tabular}{ccccccc}
       \\
       \toprule
       Properties\textbackslash Classification & Quiescent & LINERs & E+A & Seyfert & Star-Forming \\
       \midrule
       $\sigma_\ast$ (km/s)           & 213.13(54.25) & 208.12(54.73) & 210.45(86.87) & 171.20(71.16) & 99.84(52.60)\\
       $A_V$                          & -0.12(0.12) & -0.04(0.13) & -0.16(0.13) & 0.00(0.17) & 0.19(0.17)\\
       $<t_\ast>_{flux}$ (Log(Gyr))        & 9.84(0.18) & 9.85(0.16) & 9.64(0.46) & 9.49(0.47) & 8.74(0.54) \\
       $<t_\ast>_{mass}$ (Log(Gyr))        & 10.14(0.10) & 10.16(0.04) & 9.94(0.42) & 10.00(0.22) & 10.02(0.12)\\
       $<Z_\ast>_{flux}$ ($Z_\odot$)  & 1.32(0.24) & 1.28(0.27) & 1.33(0.33) & 1.33(0.33) & 0.79(0.25)\\
       $<Z_\ast>_{mass}$ ($Z_\odot$)  & 1.30(0.21) &  1.27(0.24) & 1.34(0.26) & 1.37(0.13) & 1.27(0.24)\\
       $M_{\ast,current}$ ($\log(M_\odot)$) &  11.25(0.26) & 11.12(0.40) & 11.00(0.79) & 10.98(0.43) & 10.71(0.61) \\
       \bottomrule
\end{tabular}
\end{table*}

\subsubsection{Summary for the Sample}

We summarize the stellar population properties for different types
of EGs in Table 5. The mean values and the corresponding standard
deviations of velocity dispersion, dust extinction, average age,
metallicity and the current stellar mass are listed.

From this table, we could find SFEGs have the largest difference
between the flux-weighted and mass-weighted average age, and this is
the same for the average metallicity. This demonstrates that the
ongoing star formation in SFEGs has great influence on the integral
stellar properties. At the same time, their mass-weighted
metallicities and ages are much closer to the quiescent EGs. These
SFEGs are not different objects compared with the quiescent EGs.
Essentially, they could have similar SFH that are only different in
the intensity and time-scale of the star-forming activity. It is
possible that these two types of EGs are in the same sequence of
evolution which are only affected by the mass and the environment,
we leave the detailed discussion at below.

\subsection{Absorption Line Indices}

In the study of EGs, the use of absorption lines can be traced back
to more than 20 years ago (Burstein et al. 1984). Many important
discoveries were made based on this method. For example: the famous
$Mg_2-\sigma_\ast$ relation (Kuntschner et al. 2000; Denicolo et al.
2005), where the value of $Mg_2$ index increases for more massive
EGs. This relation was considered as a mass-metallicity relation
although the age and [$\alpha/Fe$] also play roles (Thomas \&
Maraston 2003). The wildly used Lick/IDS
 system has 25 indices which can be separated into groups
for different applications. For example, the Balmer indices are good
indicators of the stellar age, especially the H$_\beta$ index, which
is very sensitive to young star formation though it is often
contaminated by nebular emission. For the metallicity, there are two
groups of elements, the $\alpha$-elements which are mainly from the
Type II supernovae and the Fe-peak elements that is associated with
the Type Ia supernovae. The indices like Mg1, Mg2, Mgb are good
tracers of $\alpha$-elements and the Fe-peak elements are often
represented by Fe5270, Fe5335 etc.. Using these indices, the stellar
population properties could be obtained. Here we present a simple
analysis using Lick absorption indices, which are taken from the
MPA/JHU DR4 VAGC database.

\begin{figure*}
  \includegraphics[width=16cm]{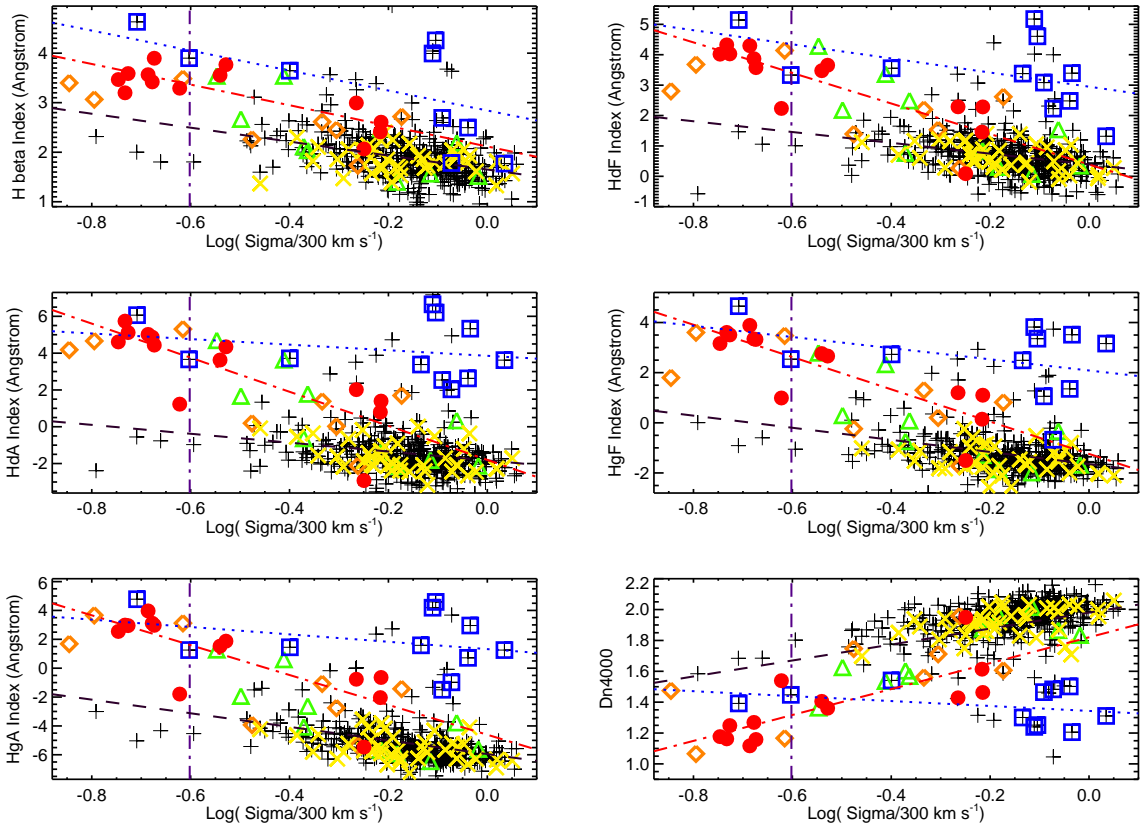}
   \caption{The first group of correlation between absorption line indices and the velocity dispersion. The indices here include
H$_\beta$, $H_{\delta}A$, $H_{\delta}F$, $H_{\gamma}A$,
$H_{\gamma}F$, $D_n4000$. }
\end{figure*}

\begin{figure*}
   \includegraphics[width=16cm]{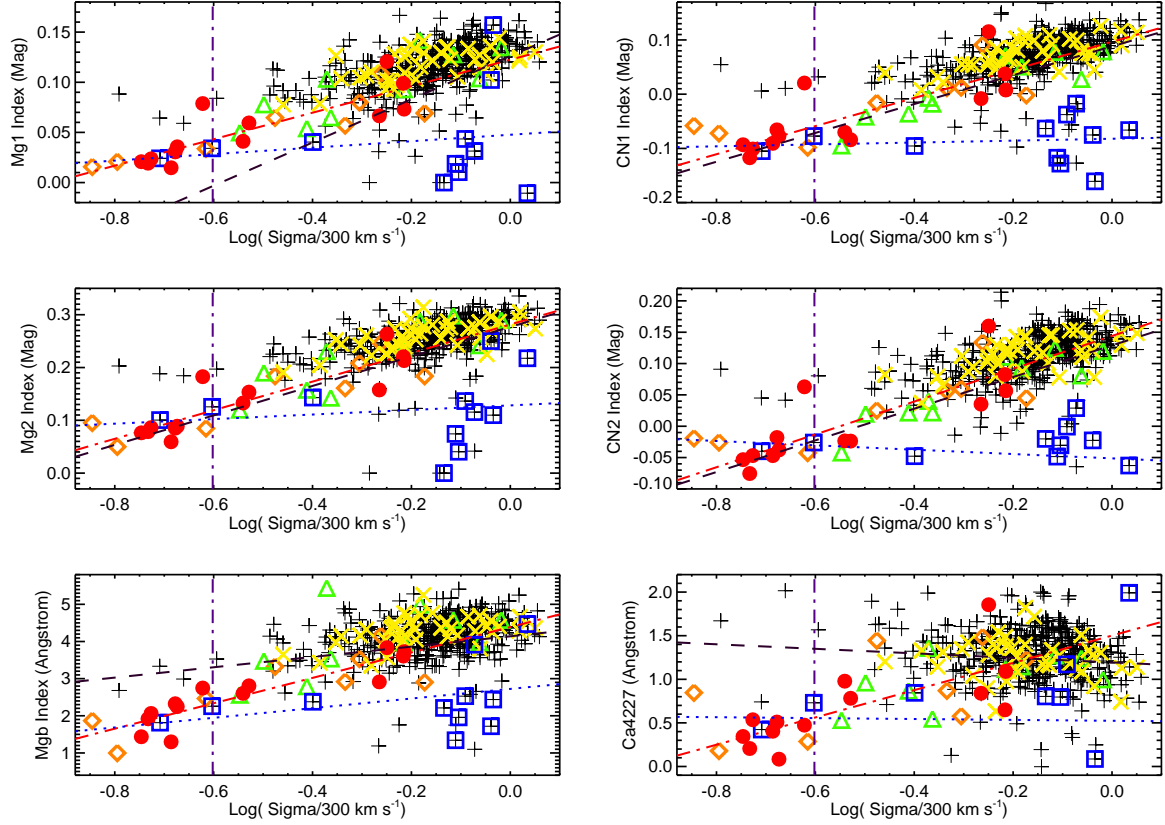}
   \caption{The second group of correlation between absorption line indices and the velocity dispersion. The indices here include
Mg1, Mg2, $Mg_bF$, CN1, CN2, Ca4227. }
\end{figure*}

\begin{figure*}
   \includegraphics[width=16cm]{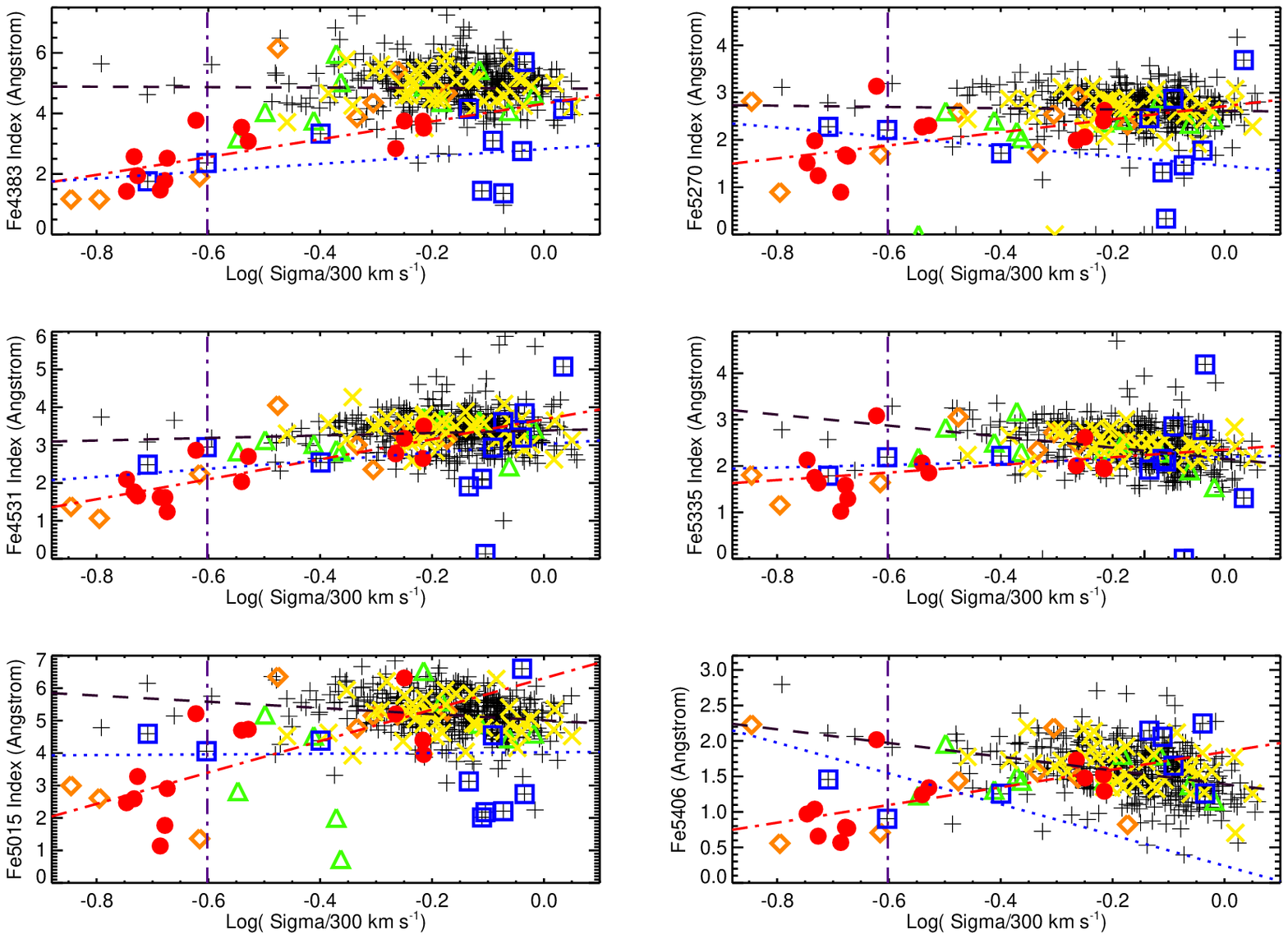}
   \caption{The third group of correlation between absorption line indices and the velocity dispersion. The indices here include
Fe4383, Fe4531, Fe5015, Fe5270, Fe5335, Fe5406. }
\end{figure*}

\subsubsection{Correlation With Galaxy Mass}

Many authors mentioned different types of $\sigma$-index relations
(mass-index relations)(Gonzalez 1993; Kuntschner et al. 2000; Nelan
et al. 2005; Annibali et al. 2007; Trager, Faber \& Dressler 2008).
In Figs 15, 16 and 17, we plot the correlation of velocity
dispersion against 18 different indices, which are separated into
three groups and the linear fits of the relations for quiescent, E+A
and star-forming EGs are also plotted as dashed lines.

The indices in Fig.15 are mostly Balmer indices, which could be used
as the age indicator. The first five indices are H$_\beta$,
H$_{\delta}A$, H$_{\delta}F$, H$_{\gamma}A$, H$_{\gamma}F$ and
negative correlation with velocity dispersion is found for each of
them. We find that the stellar age in EGs decreases as the mass
increases, which is consistent with the mass-age relation from
\texttt{STARLIGHT} fitting. For H$_{\delta}A$, H$_{\delta}F$,
H$_{\gamma}A$ and H$_{\gamma}F$, except for the E+A EGs sample, two
types of relations exit, the quiescent EGs and LINERs have a
relative flat slope, while the slopes of SFEGs, Seyfert and
transition region EGs are much steeper. These indices, especially
the H$_\gamma A$, are thought to be more sensitive to the younger
stellar population, so the difference may indicate that their
general stellar properties are different. Besides that, a strange
distribution of E+A EGs is found in these plots. For the properties
mentioned before, the E+A EGs are usually closer to the quiescent
EGs on average, but, from these age-sensitive indices, they become
much closer to the SFEGs. This should not be unexpected if we
consider the criteria used for the selection of E+A EGs sample and
it can be explained by the character of post-starburst. In the last
plot, we use the D$_n$4000 index (Balogh et al. 1999) which is also
a age indicator but more sensitive to the old stellar populations.
The positive correlation of D$_n$4000 actually has the same meaning.
These $\sigma_\ast$-index relations here confirm the results that
the stellar population in E+A EGs have measurable difference with
quiescent EGs.

In Fig.16, we concentrate on the indices that are associated with
$\alpha$-elements. The clear positive correlations are presented for
almost every index except for the Ca4227. The correlations for
different types of EGs are almost the same except for the E+A EGs
which show obvious separation from quiescent EGs again. If these
indices are connected with the $\alpha$-elements abundance in EGs,
these correlations can be seen as a mass-metallicity relation which
means more massive EGs tend to have higher $\alpha$-elements
abundance (Annibali et al. 2006; 2007), resulted from shorter
time-scale of early star-formation and higher efficiency of chemical
evolution process . From the average metallicity obtained by spectra
fitting, no correlation between mass and metallicity is found. But
it is obvious from the $\alpha$-elements sensitive indices, this
difference may reflect that there are different kinds of elements
influence the total metallicity. The famous Mg$_2$-$\sigma_\ast$
relation for our sample is fitted as:

\begin{equation}
   Mg_2=-0.375(\pm0.29)+0.264(\pm0.128)\log(\sigma_\ast)
\end{equation}

We also check the correlation of mass and Fe-peak elements sensitive
indices like Fe5270 and Fe5335. But we don't find  any significant correlation.
This is very different situation when compared to the
correlations for $\alpha$-elements. The formation of Fe-peak
elements is associated with the Type Ia supernovae, the process is
different from the $\alpha$-elements and may be not affected by the
galaxy mass. Though the correlation for the entire sample is
unconspicuous, there is still evidence which indicates the SFEGs are
also poor in Fe-peak elements when comparing with the quiescent EGs.
This could be seen as another proof that the SFEGs have different
SFH from quiescent EGs. In the plots of age and $\alpha$-element
sensitive indices, the E+A galaxies with high velocity dispersion
show very clear separation with the quiescent EGs, but this
difference is much less in the Fe-peak elements sensitive indices.

\begin{figure}
  \includegraphics[width=9.2cm]{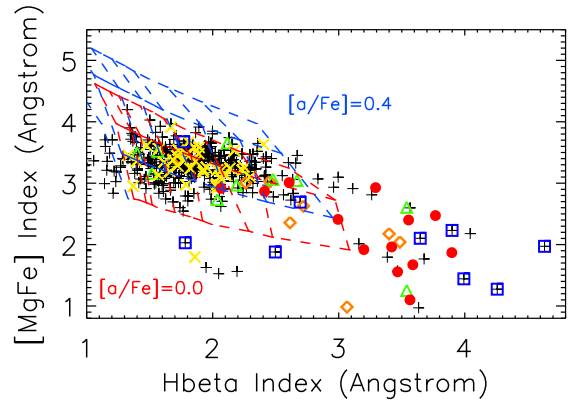}
   \caption{The absorption diagnostic diagram of H$_\beta$ and [MgFe] indices. }
\end{figure}

\begin{figure}
  \includegraphics[width=9.2cm]{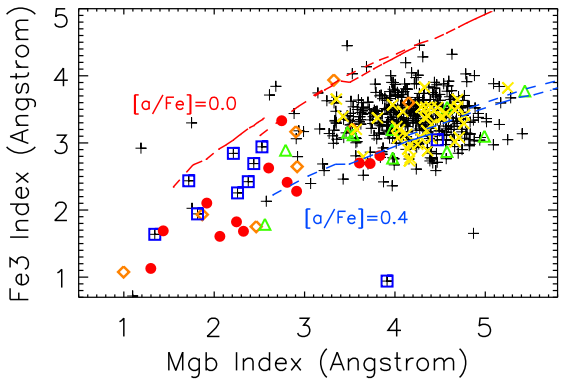}
   \caption{The absorption diagnostic diagram of $Mg_b$ and $<Fe3>$ indices. }
\end{figure}

\subsubsection{Absorption Line Diagnostic Diagram}

When using the absorption line indices for stellar population study,
the different indices diagrams are always effective method for
breaking the age-metallicity degeneracy and estimating the age,
metallicity and even the $\alpha$-element-to-iron abundance. There
are several models based on the Lick/IDS system, for example,
the most wide-used model by Thomas, Maraston \&
Bender (2003). But since the measurements of indices in our work are
not well calibrated into Lick/IDS system, our results can not be
compared with these models.

Instead, we use the synthetic spectral models for alpha-enhanced
stellar populations from Coelho et al. (2007)\footnote{G$\alpha$
models: http://www2.iap.fr/users/pcoelho/alphamodels.html}, which
allows us to explore the influence of change of the $[\alpha/Fe]$ on
the high-resolution spectral properties of evolving stellar
population. The model covers three different iron abundances
([Fe/H]=-0.5, 0.0, 0.2) and two $[\alpha/Fe]$ (0.0, 0.4) for stellar
populations between 3 and 14 Gyr. From these
models, we can predict absorption indices used in our work
under the resolution of SDSS spectra which are consistence with our
measurements. This model is a powerful tool for extracting
information about chemical properties of EGs.

A very useful diagnostic diagram is constituted by an age-sensitive
plus a metallicity sensitive indices. The H$_\beta$ index is very
good age indicator, especially for the young stellar population
involving in the recent star formation activity (Tantalo et al.
2004). The situation for metallicity is somewhat more complicated,
we already notice the different pattern for $\alpha$-elements and
Fe-peak elements. So, we decide to use the so-called [MgFe] index
which is almost free from the $\alpha$-enhanced effect, this element
is defined as $[MgFe]=\sqrt{Mgb\times(0.72Fe5270+0.28Fe5335)}$ here
(Thomas, Maraston \& Bender 2003). The H$_\beta$-[MgFe] plot can be
found at Fig.18 where we also plot the models grids we used on the
figure, the red grid is for the models with $[\alpha/Fe]=0.0$ and
the blue grid is for $[\alpha/Fe]=0.4$. From left to right, the age
is decreasing from 14 Gyr to 4 Gyr and from the bottom to top, the
metallicity is increasing from $[Fe/H]=-0.5$ to 0.2. In this figure,
the models grids are covered the distribution of most EGs in our
sample, except for a small fraction of EGs which are mainly from the
SFEGs and E+A EGs sub-sample. These galaxies are left out from the
models in the younger age and lower metallicity end, the trend is
consistent with what we got from \texttt{STARLIGHT} fitting.
Generally, the models will give much younger age estimate for most
EGs in our sample, even younger than the estimate from flux-weighted
average age. This is because the age indicator we used here, the
H$_\beta$ index, is very sensitive to the active star-formation and
young stellar population formed in the recent star-formation. And
this is also the reason why the E+A EGs can have much lower
age-estimate even younger than SFEGs while their age-estimated from
\texttt{STARLIGHT} are pretty close to the quiescent EGs.

\begin{figure}
   \includegraphics[width=9.2cm]{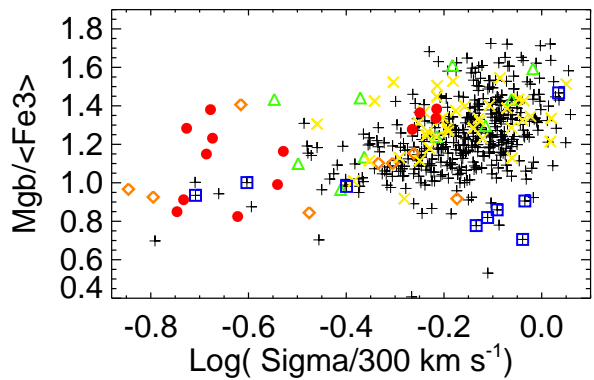}
   \caption{The correlation of the $Mg_b/<Fe3>$ ratio with the velocity dispersion estimate. }
\end{figure}

The $\alpha$-enhanced effect is thought to be the key issue in the
chemical evolution of EGs with different mass or other properties.
This effect concern the star-formation time scale, the efficiency of
chemical evolution and maybe indicate a different IMF. In our work,
we do not give a direct estimate of the $[\alpha/Fe]$ effect, but
still we can qualitatively check with the absorption line diagnostic
diagram. In Fig.19, we plot another diagram for examining the
$[\alpha]$-enhanced effect in our EGs sample, the two index we use
are the $Mg_b$ index which represent the abundance of
$\alpha$-elements and the $<Fe3>$ index which defined as
$<Fe3>=(Fe5015+Fe5270+Fe5335)/3$ for the abundance of Fe-peak
elements (Kuntschner 2000). Since both indices are metallicity
sensitive, the models grids almost degrade into a model "line" in
the figure. The models with different $[\alpha/Fe]$ are the same in
different color.

The most important result from this figure is that we can see the
$\alpha$-elements enhanced effect is really common for the EGs in
our sample, though we do not have the detailed estimate of
$[\alpha/Fe]$. In some works, there has been reported the
$\alpha$-enhanced effect is increasing with the galaxy mass
(Worthey, Faber \& Gonzalez 1992, Trager et al. 2000, Thomas et al.
2005, Nelan et al. 2005). For our sample, this is not very clear.
More directly, we can use the ratio of $Mg_b$ and $<Fe3>$ as a very
rough proxy of the $[\alpha/Fe]$ ratio. We plot its correlation with
galaxy mass estimate in Fig.20 and we find the relation maybe exists
for the quiescent EGs though the trend is still unclear. For the
entire sample, we find no significant correlation.

At the end of this section, a statistical comparison of some useful
indices of different types of EGs is shown in Table 6.

\begin{table*}
\caption{The average value and corresponding standard deviation of
some important absorption line indices for different types of
elliptical galaxies} \label{tab:6} \centering
\begin{tabular}{ccccccc}
       \\
       \toprule
       Properties\textbackslash Classification & Quiescent & LINERs & Seyfert & E+A & Star-Forming \\
       \midrule
       \Hb  & 1.88(0.79)   & 1.85(0.26)   & 2.31(0.77)   & 3.26(1.77)   & 3.12(0.64) \\
       Mg$_b$  & 3.91(1.13)   & 4.17(0.78)   & 3.99(0.93) & 2.46(0.93) & 2.58(0.78) \\
       Fe3 & 3.28(0.78) & 3.30(0.28) & 3.02(0.53) & 2.12(0.89) & 2.14(0.66)\\
       $[MgFe]'$  & 2.24(0.42) & 3.20(0.61) & 2.95(0.69) & 2.16(0.67) & 2.28(0.61)\\
       \bottomrule
\end{tabular}
\end{table*}

\section{Discussion}

In this work, we show that SFEGs deviate from the normal or
quiescent EGs, we try to explain it as a certain stage of evolution
sequence of EGs that surmised from the similarity in their SFHs and
some other aspects. Though this hypothesis and the two-component SFH
is pretty consistent with our results, it is over simple for the
theory of EGs evolution and many problems remain
unsolved. Here, we present a brief discussion about several important
ones, like the triggering mechanisms of the secondary starburst, the
degree of confirmation for the evolution sequence and the EGs
formation scenario.

\subsection{Triggering Mechanisms}

In this work, we find 13 EGs with ongoing star-forming activity and
11 post-starburst EGs, \textbf{which show the generality of the
secondary star formation activity in local EGs. This phenomenon has
already been reported (Yi et al. 2005; Jeong et al. 2007). In fact,
the optical band is not perfect for searching the evidence of
residual star-formation in EGs, the result from GALEX already
pointed out that there are much more EGs in the local universe than
we thought have different level of residual or recent
star-formation.}

Since the secondary star formation activity is considered as a
rejuvenation phenomena, the most possible origin might be the
galaxy-galaxy merger or interaction. However, in our sample, we do
not find any obvious evidence of morphological disturbance which
often indicates the merger and interaction. This is partly because
the spatial resolution and the depth of the image from SDSS is not
enough for detecting the morphological perturbations. Also, merger
events which are responsible for the star formation activity is
relative minor, the morphological perturbations already disappear
and the star formation activities are mostly shut down by the AGN
feed back or the depletion of cold gas, but in the low-mass EGs, the
star formation activity could continue much longer as we see in our
sample.

\subsection{Evolutionary Sequence ?}

To put all types of EGs into a uniform evolutionary sequence is a
really tempting target and the evidence for the existence of such a
sequence is already seen from our result.

From the properties we summarized before, we find that many properties show
a sequence for different types of EGS. The number of Seyfert, LINER
and composite EGs is quite small, so the average may suffer from
larger uncertainties, but we can see their connections and
relations in several observational or stellar population's
properties. For example, the optical color like u-r, the mass
estimated by velocity dispersion and the average age of their
stellar populations. All these information show us the possibility
that there could be a sequence between these different types of EGs.

This sequence should be seen from two aspects, a time-related
sequence and a mass-related sequence. The time-related sequence
means there could be an evolutionary link between the star-forming
and the quiescent EGS, may be like: Star-Forming EGs--Composite
EGs--Seyfert Galaxies--LINER--Quiescent EGs. Actually, in S07, the
author found a sequence in the same form from a much larger
early-type galaxies sample which is also selected from SDSS data,
although their method for decoding the SFH is quite different from
what we used here. This sequence is also supported by the concept of
AGN feedback (Ciotti \& Ostriker 1997; Silk \& Rees 1998) who is
responsible for shutting down of star-formation activity.

By the stellar population synthesis method we used here, the SFHs of
different types of EGs are actually oversimple and very coarse, we
can not ascertain the accurate age and the intensity of the
secondary starburst. So the evolutionary sequence we proposed above
is generally a logical guess. Based on this hypothesis, we see
clearly another sequence which is dominated by the mass of the
galaxies which means the more massive of the EGs, the more rapid
they evolve. The $alpha$-elements enhancement in the massive EGs
suggest that the star formation activities were more efficient and
rapid than their low-mass counterparts. The time scale of the
secondary starburst estimated by the \texttt{STARLIGHT} supports the
result too.

\textbf{Besides the galaxy property like stellar mass, the influence
from environment should be considered too. Although our sample is
not suitable for the discussion of environment (relative small and
only from a small area in the sky), we found among the 13 SFEGs,
none of them is in the cluster, which at least show that the
low-density environment may be more suitable for the residual star
formation. }

\textbf{We already knew that galaxies in different stage of merger
could show different level of activity and make up an evolutionary
sequence. But, since we excluded the galaxies with obvious disturbed
morphology, none of the EGs in our sample is in the process of
strong interaction or major merger, which means that it is hard to
determine the existence of such a merger-driven evolutionary
sequence for our sample. If more accurate SFH could be derived from
these EGs, we can compare them to the SFH of galaxies that are in
the stage of on-going merger, then maybe the relations between
different types of EGs with the mergers could be found.}

Another interesting question is if the evolutionary sequence is
really exist, where should the E+A EGs be placed. The stellar
population properties show their similarity with quiescent and
star-forming EGs in different way, we believe these galaxies are
post-starburst galaxies, but their role in the evolution of EGs is
still unclear. Also we should remember, if the star-formation
activity is triggered by merger events, this sequence of evolution
could happen more than once in the history of EGs, the real SFH
could show evidence of impacts by several evolutionary sequences
though the SFH we recovered didn't have enough resolution. And for
the E+A EGs in this sample, maybe it is not appropriate to put them
about 1 Gyrs behind the SFEGs on the evolutionary sequence. From the
mass fraction of young and intermediate age stellar population in
E+A EGs and SFEGs, it seems like that the E+A EGs had more active
star formation than the SFEGs. The SFH of E+A EGs is actually an
interesting question to be answered.

\section{Conclusions}

We have studied the properties of EGs with recent star-forming activity in
the local universe. Our main purpose is to establish the differences as well as the possible evolutionary connections among
the different types of EGs, and thus to place star-forming EGs (SFEGs) in the correct sequence of EGs' evolution.

The main results of this paper are the  following:

\begin{enumerate}

    \item  From a sample of 487 local ((z$<0.16$) EGs drawn from the SDSS DR3  we discover 13 galaxies with obvious active
star-forming activity from emission line diagnostic diagrams, and 11 E+A post-starburst galaxies. Among the 13 SFEGs, 7
have unambiguous morphological classifications as EGs while the remaining 6 galaxies have been classified as $T=0.5$. From a visual inspection of their
images we found no obvious difference between the morphology of these 6 galaxies from those of typical EGs. The fraction of SFEGs is $2.7\%$ (or $1.4\%$
when consider only the 7 ones with $T=0$) and the fraction for E+As is $2.3\%$.

 \item SFEGs  shows some obvious broad differences from  quiescent EGs:  smaller sizes, lower masses,  and lower luminosities. Their optical colors are
bluer than those of normal EGs.  In particular, however, we actually found  four
SFEGs that have  luminosities and masses  typical of quiescent EGs. On the other hand, our E+As show a remarkable  similarity in mass, luminosity, and color to the quiescent EGs.

 \item From stellar population synthesis (\texttt{STARLIGHT})
fitting, we estimated the distribution of different stellar
populations. We found for the local EGs in our sample, the star-formation history can
be described by a minor "juvenile" stellar population formed during  secondary star-forming activity around 1Gyr ago on the
background of a dominant old  stellar population  formed a long time ago during a very intense period of star formation. This simple
description is basically valid for all types of EGs including the quiescent EGs that show no evidence of recent star-formation at
all. From this point of view we can say the secondary star-forming activity, the so-called "juvenile" effect, is quite common in local
EGs.

 \item We found a mass-age relation indicating that more massive
EGs tend to have older average ages.  A similar correlation between age and metallicity, however, was not found.
From population synthesis we find that our SFEGs  have younger ages (especially from the flux-weighted average age) and
lower average metallicities. Also,  due to the active star-formation these EGs  have higher dust extinction. The discrepancy between mass estimates from stellar populations and from the velocity dispersion may be due to the presence of a central stellar disc or some other rotationally supported component associated with a recent star-formation event, and which can lower the velocity dispersion estimate. 

 \item From our simple analysis of  absorption line indices  we find several interesting results. First, from our analysis of the
$\sigma_\ast$-index relation, we found a good correlation between age and $\alpha$-element sensitive indices, but no relation was found
for Iron-peak sensitive indices. These results confirm the mass-age relation and suggest that the $\alpha$-element abundances tend to be higher in more massive galaxies, but the Fe-peak elements abundance seem to show no correlation with galaxy mass.  Second, we notice  that E+As, which have masses (velocity dispersions) similar to those of quiescent EGs,  show completely different  distributions in most of the $\sigma_\ast$-index
diagrams, and specially in the relation between age and $\alpha$-element sensitive
indices. In fact, line indices of E+A galaxies are much closer to those of SFEGs than to normal ellipticals. Third, we found that the $\alpha$-enhanced effect is very common
among local EGs, but we do not find a correlation with galaxy mass.

 \item From the different properties discussed above we consider that SFEGs
are not a particularly special type of EGs: they are not "young" objects since
the main stellar populations are as old as those of normal EGs while
their star formation histories show considerable similarity with the quiescent ones.
This seems to indicate that SFEGs are just passing through an evolutionary stage, which is more pronounced  for low mass EGs, but is still part of a common evolutionary sequence for Elliptical galaxies that is determined by total mass and environment.

\end{enumerate}

\section*{Acknowledgments}
The authors are very grateful to the anonymous referee for his/her
thoughtful and instructive comments that significantly improved the
content of the paper. We thank Kevin Schawinski, Cid Fernandes,
Daniel Thomas, Harald, Kuntschner, Paula Coelho,  Anna Gallazzi,
Jarle Brinchmann, Yanbin Yang  and Jorge Melnick for helpful
comments and suggestions.

This work is supported by Program for New Century Excellent Talents
in University (NCET), the National Natural Science Foundation of
China under grants (10878010, 10221001, and10633040), and  the
National Basic Research Program (973 program No. 2007CB815405).  The
STARLIGHT project is supported by the Brazilian agencies CNPq, CAPES
and FAPESP and by the France-Brazil CAPES/Cofecub program.

We acknowledge use of the public data of the MPA/JHU Garching DR4
data, and much of the exploratory work that led to the results
contained in this paper benefited from an SQL data base created with
William Schoenell as part of the SEAGal project

Funding for the creation and distribution of the SDSS Archive has
been provided by the Alfred P. Sloan Foundation, the Participating
Institutions, the National Aeronautics and Space Administration, the
National Science Foundation, the US Department of Energy, the
Japanese Monbukagakusho, and the Max Planck Society. The SDSS Web
site is http://www.sdss.org.

The SDSS is managed by the Astrophysical Research Consortium (ARC)
for the Participating Institutions. The Participating Institutions
are The University of Chicago, Fermilab, the Institute for Advanced
Study, the Japan Participation Group, The Johns Hopkins University,
Los Alamos National Laboratory, the Max- Planck-Institute for
Astronomy (MPIA), the Max-Planck-Institute for Astrophysics (MPA),
New Mexico State University, University of Pittsburgh, Princeton
University, the United States Naval Observatory and the University
of Washington.

\bsp

\label{lastpage}

\end{document}